\documentclass[11pt,a4paper,english,twoside]{article}

\usepackage{a4wide}
\usepackage{amssymb, amsmath, amsthm}
\usepackage{graphicx}
\usepackage{subcaption}
\usepackage[all]{xy}
\usepackage{enumerate}
\usepackage[pdftex,hyperref,svgnames]{xcolor}
\usepackage[pdftex,colorlinks=true,
pdfstartview=FitV,
pdfnewwindow=true,
linktoc = page,
linkcolor= blue,
citecolor= red,
urlcolor= blue,
hyperindex=true,
hyperfigures=false]{hyperref}
\hypersetup{linktocpage}
\usepackage{dsfont}
\usepackage{empheq}
\usepackage{cite}
\usepackage{float}
\usepackage{cancel}
\usepackage{relsize}
\usepackage{soul}
\usepackage{enumitem}
\usepackage{hhline}
\usepackage{multirow}

\newcommand{\beq}{\begin{equation}}
\newcommand{\eeq}{\end{equation}}
\def\bea#1\eea{\begin{align}#1\end{align}}
\def\beal#1\eeal{\begin{subequations}\begin{align}#1\end{align}\end{subequations}}
\newcommand{\nn}{\nonumber}
\newcommand{\w}{\wedge}
\newcommand{\R}{\mathcal{R}}

\def\del {\partial}
\def\d {{\rm d}}
\def\mmm {\mathcal{M}_{10-d}}

\begin{document}
\numberwithin{equation}{section}

\begin{titlepage}

\begin{center}

\phantom{DRAFT}

\vspace{1.2cm}

{\LARGE \bf{(Quasi-) de Sitter solutions across dimensions\vspace{0.4cm}\\ and the TCC bound}}\\

\vspace{2.2 cm} {\Large David Andriot$^{1}$, Ludwig Horer$^{1, 2}$}\\
\vspace{0.9 cm} {\small\slshape $^1$ Laboratoire d’Annecy-le-Vieux de Physique Th\'eorique (LAPTh),\\
CNRS, Universit\'e Savoie Mont Blanc (USMB), UMR 5108,\\
9 Chemin de Bellevue, 74940 Annecy, France}\\
\vspace{0.2 cm} {\small\slshape $^2$ Institute for Theoretical Physics, TU Wien\\
Wiedner Hauptstrasse 8-10/136, A-1040 Vienna, Austria}\\
\vspace{0.5cm} {\upshape\ttfamily andriot@lapth.cnrs.fr; ludwig.horer@tuwien.ac.at}\\

\vspace{2.8cm}

{\bf Abstract}
\vspace{0.1cm}
\end{center}

\begin{quotation}
In this work, we investigate the existence of string theory solutions with a $d$-dimensional (quasi-) de Sitter spacetime, for $3 \leq d \leq 10$. Considering classical compactifications, we derive no-go theorems valid for general $d$. We use them to exclude (quasi-) de Sitter solutions for $d \geq 7$. In addition, such solutions are found unlikely to exist in $d=6,5$. For each no-go theorem, we further compute the $d$-dependent parameter $c$ of the swampland de Sitter conjecture, $M_p \frac{|\nabla V|}{V} \geq c$. Remarkably, the TCC bound $c \geq \frac{2}{\sqrt{(d-1)(d-2)}}$ is then perfectly satisfied for $d \geq 4$, with several saturation cases. However, we observe a violation of this bound in $d=3$. We finally comment on related proposals in the literature, on the swampland distance conjecture and its decay rate, and on the so-called accelerated expansion bound.
\end{quotation}

\end{titlepage}

\newpage

\tableofcontents

\section{Introduction and results summary}

While current observations indicate that we live in a $4$-dimensional (4d) spacetime, string theory, as a candidate for a fundamental theory of nature, rather advocates 10 dimensions. A standard scenario to accommodate this situation is to consider the corresponding 6 extra space dimensions to be compact and small, allowing them to be so far undetected. From the string theory perspective, it remains a natural question to ask: why 4 dimensions \cite{Brandenberger:1988aj}? In this paper, we will consider a $d$-dimensional extended spacetime, $3 \leq d \leq 10$, and $10-d$ extra dimensions gathered in a compact manifold $\mmm$, and we will provide first hints at a preference for $d=4$.

Beyond its dimension, another important characteristic of our spacetime is its geometry: our universe is currently observed to be in accelerated expansion, and this is also suspected to have happened in its early days, during a so-called inflation phase. A possibly corresponding geometry would be a $d$-dimensional de Sitter spacetime, for us with $d=4$, with a cosmological constant $\Lambda_d$ responsible for the accelerated expansion. Another option is a slight deviation thereof, namely a quasi-de Sitter spacetime, that we now explain. A standard effective cosmological model in $d$ dimensions, $d\geq 3$, is given as follows. It involves gravity and minimally coupled scalar fields $\varphi^i$, subject to a scalar potential $V$, as described by the following action
\beq
{\cal S}= \int \d^d x \sqrt{|g_d|} \left(\frac{M_p^2}{2} \R_d - \frac{1}{2} g_{ij} \del_{\mu}\varphi^i \del^{\mu}\varphi^j - V \right) \ , \label{action}
\eeq
with the reduced $d$-dimensional Planck mass $M_p$, and the field space metric $g_{ij}$. A solution to this model with a de Sitter spacetime is a critical point of the potential, $\nabla V=0$, such that $\varphi^i$ have no kinetic energy, and $\Lambda_d = \frac{V}{M_p^2} = \frac{d-2}{2d} \R_d > 0$. A quasi-de Sitter solution is a slight deviation: $V >0$, the slope of the potential $|\nabla V|$ is small, and fields are rolling. Is it possible to obtain such an effective model with such solutions from string theory? For our universe, we are interested in $d=4$, but since there is a priori no preference for this dimension, we investigate in this paper the possibility of finding (quasi-) de Sitter solutions across dimensions. Such an investigation is actually part of the swampland program \cite{Vafa:2005ui, Palti:2019pca}, which aims at characterising all that can be obtained from string theory. From the swampland perspective, all dimensions $d$, with $3 \leq d \leq 10$,\footnote{Note that $d=3$ is nevertheless debated, as discussed in Section \ref{sec:cvalues}.} should be treated on equal footing, and one is then entitled to ask whether (quasi-) de Sitter solutions can be obtained from string theory for any dimension $d$.

A partial answer is that it is notoriously difficult to find such solutions in a well-controlled model from string theory. In the swampland program, this has a led to propose the so-called de Sitter conjecture \cite{Obied:2018sgi, Andriot:2018wzk, Garg:2018reu, Ooguri:2018wrx, Andriot:2018mav, Rudelius:2019cfh, Bedroya:2019snp}: it claims a systematic obstruction to such solutions (in models \eqref{action} coming from string theory) in the form of an inequality
\beq
|\nabla V| \geq \frac{c}{M_p}\, V \ , \label{nablaVintro}
\eeq
where $c \sim \mathcal{O}(1)$. Indeed, $|\nabla V|$ obeying \eqref{nablaVintro} cannot vanish, and cannot even be small. Although mostly tested in $d=4$ \cite{Andriot:2020lea}, together with the difficulties met by constructions of $d=4$ de Sitter solutions, the conjecture is a priori valid in all $d \geq 3$. The inequality \eqref{nablaVintro} is nowadays believed to hold only in the asymptotics of field space. It is for instance the case for the most refined version of this conjecture, the Trans-Planckian Censorship Conjecture (TCC) \cite{Bedroya:2019snp}, which in addition provides the following $d$-dependent lower bound on $c$
\beq
\text{TCC bound:} \qquad c\geq  c_0 = \frac{2}{\sqrt{(d-1)(d-2)}} \ . \label{TCCbound}
\eeq
In this paper, we will test this quantitative bound thanks to a study in all dimensions $d \geq 3$. Note that if this bound had to be obeyed in our universe, much more involved cosmological models would then be required \cite{Agrawal:2018own, Bedroya:2019tba, Agrawal:2020xek, Rudelius:2022gbz}.\\

We investigate in this work the possibility of finding solutions with a $d$-dimensional (quasi-) de Sitter spacetime, in compactifications of 10d string theory. We do so restricting to classical string backgrounds: this simple setting is easier to control, compared to other approaches requiring perturbative or non-perturbative contributions \cite{Kachru:2003aw, Balasubramanian:2005zx, Gao:2020xqh, Junghans:2022exo, Junghans:2022kxg}. Let us report already that we will observe many constraints or even exclusions of quasi-de Sitter solutions in higher dimensions $d>4$, and other, non-classical approaches may as well face related difficulties; in particular, we are not aware of any attempt to obtain such solutions in $d>4$ using these other approaches. It would be interesting to extend the present analysis to these less perturbative settings. However, another motivation for focusing on the classical string regime is that it should correspond to the asymptotics of the effective field space, where the inequality \eqref{nablaVintro} is thought to be valid. It is thus the right setting to test the de Sitter swampland conjecture and the TCC, as we will do.

Classical string backgrounds with a $d$-dimensional de Sitter spacetime, in short classical de Sitter solutions, are usually searched for in compactifications of 10d type IIA/B supergravities. Forbidding a de Sitter solution in supergravity, as we will do here with the derivation of no-go theorems, is sufficient to exclude a classical de Sitter string background. The converse is however not true: whenever a de Sitter solution of 10d supergravity is found, as recently done most extensively in \cite{Andriot:2022way,Andriot:2022yyj} for $d=4$, one still has to verify that the solution is in the classical string regime, justifying the approximation to 10d supergravity. This last check turns out to fail in the $d=4$ examples where it has been attempted \cite{Banlaki:2018ayh, Andriot:2020vlg}. While general arguments for such failure (in $d=4$) have been put forward \cite{Junghans:2018gdb, Andriot:2019wrs, Grimm:2019ixq}, consistently with the de Sitter conjecture, it remains an open issue whether classical de Sitter solutions exist.

Together with type IIA/B supergravities, we allow here for $D_p$-branes and orientifold $O_p$-planes, collectively called sources. In a classical compactification, one could in principle allow for other objects, namely $NS_5$-branes or Kaluza--Klein monopoles, as well as anti-$D_p$-branes. We do not include those for various reasons, some being presented in the Introduction of \cite{Andriot:2022yyj}, the main reason being simplicity and control. Let us add that while $O_p$ are typically introduced to circumvent the Maldacena-Nu\~nez no-go theorem \cite{Maldacena:2000mw}, having together anti-$D_p$ would break supersymmetry in the effective theory, which might be phenomenologically undesired.\footnote{Another technical point is that the contribution of $O_p/D_p$ is the same in Bianchi identities and Einstein equations, two 10d equations used in establishing no-go theorems, while it is not the case for $O_p$ and anti-$D_p$, making then the analysis technically more involved.} In any case, it would be interesting to study whether the constraints obtained here with only $O_p/D_p$, in particular in higher dimensions $d>4$, can be extended when including other possible objects. Another restriction of our analysis is that we consider smeared $O_p/D_p$, going together with having a constant background dilaton and no warp factor: we believe however that it should be possible to extend our results beyond this (common) ansatz \cite{Andriot:2016xvq}. Note that our compactification is not more restricted (see Section \ref{sec:conventions}): in particular, we allow for ($d$-dimensional) spacetime-filling fluxes, fluxes are not taken constant and we do not impose any restriction on the compact manifold $\mmm$.\\

In this context, there exist many works on classical de Sitter solutions, either searching for them or constraining them, and we refer to \cite{Andriot:2019wrs} for a review, and \cite{Andriot:2021rdy} for a recent attempted exhaustive reference list.\footnote{Let us also mention the recent work \cite{Farotti:2022xsd} which considers de Sitter solutions in higher dimensions, allowing however for non-compact extra dimensions.} Given the difficulties in finding these solutions, many no-go theorems have been established, reviewed e.g.~in \cite{Andriot:2018ept, Andriot:2019wrs, Andriot:2020lea}, constraining the necessary manifold properties, the fluxes and $O_p/D_p$ contents, leading eventually to successful supergravity configurations listed in \cite{Andriot:2022way}. No-go theorems can be established in two ways. A first one is to combine 10d equations of motion and Bianchi identities to reach, upon some assumption, an inequality ${\cal R}_d \leq 0$: this forbids a de Sitter solution. A second way is to use a $d$-dimensional effective theory of the type \eqref{action}, and obtain, upon assumptions, an inequality of the type
\beq
a\, V + \sum_{i} b_i\, \varphi^i \del_{\varphi^i} V  \leq 0 \ ,\label{VdelV}
\eeq
with $a>0, \exists\, b_i \neq 0$. This forbids in the same way de Sitter critical points, and it can typically be matched with an inequality ${\cal R}_d \leq 0$ obtained in 10d. The advantage of the $d$-dimensional derivation is that the inequality \eqref{VdelV} forbids as well quasi-de Sitter solutions, as we explain in more detail in Section \ref{sec:quasidS} and \ref{sec:dnogo}. Indeed, \eqref{VdelV} can typically be rewritten in the form \eqref{nablaVintro}, allowing eventually to deduce a value for $c$. In this paper, we will extend well-known $d=4$ no-go theorems to arbitrary $d$ dimensions, $3 \leq d \leq 10$, with a few novelties related to this extension. While most results on no-go theorems are known in $d=4$, let us mention the pioneering paper \cite{VanRiet:2011yc} on such extensions to $d$ dimensions: we reproduce and extend here the results obtained there, as detailed in Section \ref{sec:dSineachd}.

We first derive no-go theorems in arbitrary $d$ using 10d equations in Section \ref{sec:10dnogo}, and then do the same in Section \ref{sec:dnogo} with an effective $d$-dimensional theory derived in Section \ref{sec:dimred}. In Section \ref{sec:dSineachd}, we make a concrete use of these no-go theorems in each dimension $d$: given that the flux and source content in higher $d$ is limited, some assumptions get automatically satisfied. This is even more true when restricting in Section \ref{sec:susy} to supersymmetry-preserving $O_p/D_p$ source configurations. As summarized in Section \ref{sec:dSineachdsum}, we eventually exclude any de Sitter solution in $d \geq 7$, leaving only a few options in $d=6,5$ (see Table \ref{tab:d=6IIAsusy}, \ref{tab:d=5IIAsusy}, \ref{tab:d=5IIBsusy}), and much more in $d=4,3$. We argue in addition why, according to Conjecture 1 of \cite{Andriot:2019wrs} and 4 of \cite{Andriot:2022way}, there should not be any such solution in $d=6,5$, hinting at $d\leq 4$ if not $d=4$. These results are straightforwardly extended to an exclusion of quasi-de Sitter solutions in Section \ref{sec:quasidS}.

The $d$-dimensional derivations of the no-go theorems in Section \ref{sec:dnogo} allow to obtain a value for $c$ associated to each no-go theorem. While such an extensive analysis had been carried out in \cite{Andriot:2020lea} in $d=4$, we obtain here a $d$-dependent $c$ value, allowing a proper comparison to the expression of the TCC bound \eqref{TCCbound}. Our results are summarized and discussed in Section \ref{sec:discussionC}. Remarkably, as displayed in Table \ref{tab:cvalues} and Figure \ref{fig:cvalues}, we obtain that the TCC bound is perfectly verified in $d \geq 4$, with several saturation cases, namely all those that already gave saturation in $d=4$. This is a non-trivial check of the TCC bound \eqref{TCCbound}, since there is a priori {\sl no reason} for this precise expression to be reproduced in all dimensions $d$ by supergravity no-go theorems. This result is however consistent with the swampland perspective and conjectures. We refer to Section \ref{sec:cvalues} for a more extensive discussion. In that section, we discuss as well an interesting result in $d=3$: a newly derived no-go theorem \eqref{eq:p2nogo} or \eqref{cd=3p=2} leads to a $c$ value that is {\sl lower} than the TCC bound. This is the only violation we know of this bound. It is probably related to peculiarities of gravity in $d=3$, as we comment on in Section \ref{sec:cvalues}. Having established the TCC bound on more solid grounds for all $d\geq 4$ with the no-go theorem derivations, we compare it in Section \ref{sec:swampconj} to other $d$-dependent proposals that appeared in the literature, in particular that of \cite{Rudelius:2021oaz}. We also discuss in Section \ref{sec:swampconj} the related swampland distance conjecture and its rate $\lambda$, more precisely tentative $d$-dependent expressions for its lower bound. We finally comment in Section \ref{sec:accexp} and Appendix \ref{ap:accexp} on an asymptotic upper bound on $\frac{|\nabla V|}{V}$, required to ensure cosmic accelerated expansion. We verify that this upper bound only holds upon some assumptions, which might as well be violated. Such a violation suggests different cosmological scenarios still allowing for accelerated expansion. Those could be particularly interesting if the TCC bound should hold true.

\section{$10$-dimensional derivations}\label{sec:10d}

We present in Section \ref{sec:conventions} the general compactification ansatz to be used, from $10$-dimensional type IIA/B supergravities towards a $d$-dimensional maximally symmetric spacetime, with $3 \leq d \leq 10$. We also provide a few 10d equations of motion. Using those, we then derive in Section \ref{sec:10dnogo} no-go theorems on the existence of solutions with a $d$-dimensional de Sitter spacetime. These no-go theorems are, most of the time, extensions of known ones in $d=4$, with a few novelties for $d=3$ or for sources of multiple dimensionalities. They will be reproduced and extended in Section \ref{sec:ddim} with a $d$-dimensional approach, forbidding in addition {\sl quasi}-de Sitter solutions, upon the same assumptions. We will use these no-go theorems in Section \ref{sec:dSineachd} to concretely constrain or even exclude (quasi-) de Sitter solutions for each dimension $d$.

\subsection{Conventions, compactification ansatz and 10d equations}\label{sec:conventions}

We start with 10-dimensional (10d) type IIA/B supergravities, together with $D_p$-branes and orientifold $O_p$-planes, collectively called sources; we follow conventions of \cite[App. A]{Andriot:2016xvq}. We consider a compactification on a $(10-d)$-dimensional compact manifold $\mmm$, to a $d$-dimensional maximally symmetric spacetime. We restrict ourselves to $d \geq 3$. We use a signature $(-, +, ...,+)$ for the $10$- and $d$-dimensional spacetimes. Let us specify the corresponding compactification ansatz of the 10d fields.

To preserve $d$-dimensional Lorentz invariance (or maximal symmetry), we consider space-filling $O_p/D_p$ sources, requiring $d-1\leq p \leq 9$. $O_p/D_p$ sources are gathered by dimensionality $p$, and for each of those in sets of parallel sources (i.e.~placed along the same dimensions) labeled by $I$. Their contributions to the equations are denoted $T_{10}^{(p)_I}$, and $T_{10}^{(p)} = \sum_I T_{10}^{(p)_I}$, $T_{10} = \sum_p T_{10}^{(p)} = T_{MN} g^{MN}$ with 10d indices $M,N$; we refer to \cite{Andriot:2019wrs, Andriot:2017jhf} for related conventions. The 10d RR fluxes are denoted $F_q^{10}$, $0\leq q \leq 5$, with the on-shell condition $F_5^{10} = - *_{10} F_5^{10}$, $|F_5^{10}|^2=0$. The 10d NSNS flux is $H^{10}$. The internal (along $\mmm$) fluxes, denoted $F_q$ and $H$, are taken to depend only on internal coordinates. Lorentz invariance in the external dimensions only allows for spacetime-filling fluxes: those are denoted $F_q^d$ and $H^d$, and are non-zero for $q \geq d$, or $d=3$ for the $H$-flux; we have $F_0^d=F_1^d=F_2^d=0$ here. One then has $F_q^{10} = F_q^d +F_q$, and similarly for $H^{10}$. Before specifying also the ansatz for the metric and dilaton, let us give more details on spacetime-filling fluxes.

By definition, $F_q^d$ for $q \geq d$ is along the $d$-dimensional volume form ${\rm vol}_d$, leaving $q-d$ legs along the $10-d$ internal dimensions. One then introduces ``fake'' internal fluxes denoted $F_{10-q}$, such that $*_{10-d} F_{10-q}$ captures these remaining internal $q-d$ legs. More precisely, the internal $F_{10-q}, H_7$ are introduced as follows
\bea
& F_q^d \equiv (-1)^{[\frac{q+1}{2}]} *_{10} F_{10-q} = (-1)^{[\frac{q+1}{2}]} (-1)^{(10-q)d}\, {\rm vol}_d \w *_{10-d} F_{10-q} \ ,\label{Fqd}\\
& H^d \equiv *_{10} H_7 = (-1)^d\, {\rm vol}_d \w *_{10-d} H_7 \ ,\nn
\eea
where the sign can be chosen freely, and follows here \cite{Andriot:2016xvq}; $[\cdot]$ denotes the integer part. This convention is consistent for $F_5$ and the anti-self duality: one verifies indeed that $F_5^{10} = - *_{10} F_5 + F_5$ satisfies $F_5^{10} = - *_{10} F_5^{10}$. For $q<d$, one has $F_q^d = F_{10-q}=0$, and for $d>3$, $H^d = H_7=0$. Finally, one gets the following squares
\beq
|F_q^{10}|^2 = |F_q|^2 - |F_{10-q}|^2 \ ,\ |H^{10}|^2 = |H|^2 - |H_7|^2 \ , \label{Fqdsquare}
\eeq
using the signature.

Let us add a word on $p=9$ sources. Using a democratic formalism, $O_9/D_9$ would be electric sources of a $C_{10}$ potential. Since the latter does not admit a field strength in 10d, it has no kinetic term, so it does not carry any propagating degree of freedom. Avoiding a singular propagator, in particular in type I string theory, then requires the sum of $O_9/D_9$ charges to vanish \cite{Polchinski:1998rr}. One can also see that the equation of motion for $C_{10}$, starting from the $O_9/D_9$ world-volume actions, requires this charge cancelation. Equivalently, the Bianchi identity sourced by $O_9/D_9$ does not contain any admissible magnetic flux, thus requiring again this cancelation,\footnote{We thank T.~Wrase for this remark.} which means
\beq
T_{10}^{(9)}=0 \ .\label{T109=0}
\eeq
This observation implies in our framework that one cannot have $O_9$ without $D_9$ and vice-versa; this will play an important role in Section \ref{sec:dSineachd}.
\\

Without more specifications, we now provide a few useful equations of motion. To start with, the dilaton equation of motion is given by (see e.g.~\cite[Sec. 6]{Andriot:2017jhf})
\beq
\label{eq:dilatoneom0}
2 \R_{10} + e^\phi \sum_{p} \frac{T^{(p)}_{10}}{p+1}-|H|^2 + |H_7|^2 +8 \left(\Delta \phi - |\del \phi|^2 \right)=0 \ .
\eeq
The 10d Einstein equations are given e.g.~in \cite[App. A]{Andriot:2016xvq}. Their trace-reversed version is given as follows for type IIA
\bea
\R_{MN} &= \frac{1}{4} H^{10}_{MPQ} {H^{10}_N}^{PQ} + \frac{e^{2\phi}}{2} \left(F_{2\ MP} {F_{2\ N}}^P + \frac{1}{3!} F^{10}_{4\ MPQR} {F^{10}_{4\ N}}^{PQR} \right) \label{eq:tracerevEqIIA} \\
& + \frac{e^\phi}{2} T_{MN} - 2 \nabla_M \del_N \phi \nn \\
& + \frac{g_{MN}}{16} \left(-e^\phi T_{10} - 2 |H|^2 + 2|H_7|^2 + e^{2 \phi} \left(|F_0|^2 - |F_2|^2 - 3 |F_4|^2 + 3 |F_6|^2\right) - 4 \Delta \phi + 8 |\del \phi|^2 \right) \nn
\eea
and for type IIB
\bea
\R_{MN} &= \frac{1}{4} H^{10}_{MPQ} {H^{10}_N}^{PQ} + \frac{e^{2\phi}}{2} \left(F_{1\ M} F_{1\ N} + \frac{1}{2!} F^{10}_{3\ MPQ} {F^{10}_{3\ N}}^{PQ} + \frac{1}{2 \cdot 4!} F^{10}_{5\ MPQRS} {F^{10}_{5\ N}}^{PQRS} \right) \nn \\
& + \frac{e^\phi}{2} T_{MN} - 2 \nabla_M \del_N \phi \label{eq:tracerevEqIIB} \\
& + \frac{g_{MN}}{16} \left(-e^\phi T_{10} - 2 |H|^2 + 2|H_7|^2 - 2 e^{2 \phi}( |F_3|^2 - |F_7|^2) - 4 \Delta \phi + 8 |\del \phi|^2 \right) \nn
\eea
where we developed the squares of 10d fluxes as explained above. The 10d Einstein trace is then
\beq
\label{eq:10deinstein0}
4 \R_{10} + \frac{e^{\phi}}{2} T_{10} - |H|^2 + |H_7|^2 - \frac{e^{2\phi}}{2} \sum_{q=0}^7 (5-q) |F_q|^2 - 20 |\del \phi|^2 + 18 \Delta \phi=0 \ .
\eeq

Let us now be more specific on our compactification ansatz for the metric and dilaton. We consider the 10d spacetime to be a direct product, in particular we consider no warp factor in the following 10d metric
\beq
\d s_{10}^2 = g_{\mu\nu}(x) \d x^{\mu} \d x^{nu} + g_{mn}(y) \d y^{m} \d y^{n} \ ,\label{metric}
\eeq
where $\mu, \nu = 0,...,d-1$, $m,n=d,...,9$. Accordingly, we consider smeared sources and a constant dilaton, with $g_s=e^\phi$. We nevertheless believe, following e.g.~\cite{Andriot:2016xvq, Cribiori:2019clo}, that the results obtained here can be generalized beyond these restrictions.

Using notations reviewed in \cite{Andriot:2019wrs}, we express the source energy momentum tensor $T_{MN}$ as $T_{AB}= {e^M}_A {e^N}_B T_{MN}$ in ``flat'' indices, using an orthonormal coframe. The metric \eqref{metric} leads to a decomposition into the $d$-dimensional $\alpha, \beta$, and the $(10-d)$-dimensional internal directions $\{a_{\parallel_I}, a_{\perp_I}\}$ parallel or transverse to each source set $I$, as follows
\bea
& T_{AB}= \delta_A^\alpha \delta_B^\beta\, T_{\alpha \beta} + \sum_{p,I} \delta_A^{a_{\parallel_I}} \delta_B^{b_{\parallel_I}}\, T^{(p)_I}_{a_{\parallel_I} b_{\parallel_I}} \\
{\rm with}\quad & T_{\alpha \beta}= \eta_{\alpha \beta} \sum_p \frac{T_{10}^{(p)}}{p+1} \ ,\quad T^{(p)_I}_{a_{\parallel_I} b_{\parallel_I}} = \delta_{a_{\parallel_I} b_{\parallel_I}} \frac{T_{10}^{(p)_I}}{p+1} \ ,\nn
\eea
while for each set $I$, $T_{a_{\perp_I} b_{\perp_I}} = {e^M}_{A_{\perp_I}} {e^N}_{B_{\perp_I}} T_{MN} = 0$. Note that we have {\sl not} restricted ourselves to group manifolds, nor considered constant fluxes.

From this ansatz, and using \eqref{eq:tracerevEqIIA}, \eqref{eq:tracerevEqIIB}, we obtain the $d$-dimensional Einstein equations
\bea
\R_{\mu \nu} &= \frac{g_{\mu \nu}}{16} \left( g_s \sum_p \frac{T_{10}^{(p)}}{p+1} (7-p) - 2 |H|^2 - 6|H_7|^2 + g_s^2 \left( |F_0|^2 - |F_2|^2 - 3 |F_4|^2 - 5 |F_6|^2\right) \right)\nn\\
\R_{\mu \nu} &= \frac{g_{\mu \nu}}{16} \left( g_s \sum_p \frac{T_{10}^{(p)}}{p+1} (7-p) - 2 |H|^2 - 6|H_7|^2 - g_s^2 \left( 2 |F_3|^2 + 4 |F_5|^2 + 6 |F_7|^2 \right) \right)\label{eq:ddimEq}
\eea
where we used that $1/(q-1)!\  F^{10}_{q\ \mu P...Q} {F^{10}_{q\ \nu}}^{P...Q} = - g_{\mu\nu} |F_{10-q}|^2$.\\

We deduce the $d$-dimensional Einstein trace, that we give together with the dilaton equation of motion \eqref{eq:dilatoneom0} and the 10d Einstein trace \eqref{eq:10deinstein0} for our ansatz in type IIA/B

\begin{empheq}[box=\boxed]{align}
& 2 \R_{10} + g_s \sum_{p} \frac{T^{(p)}_{10}}{p+1}-|H|^2 + |H_7|^2 =0 \label{eq:dilatoneom} \\
& 4 \R_{10} + \frac{g_s}{2} T_{10} - |H|^2 + |H_7|^2 - \frac{g_s^2}{2} \sum_{q=0}^7 (5-q) |F_q|^2 =0 \label{eq:10deinstein}\\
& \R_d= \frac{d}{16} \left( g_s \sum_p \frac{7-p}{p+1}\, T_{10}^{(p)} - 2|H|^2 - 6|H_7|^2 + g_s^2 \sum_{q=0}^7 (1-q) |F_q|^2\right) \label{eq:dRicciScalar}
\end{empheq}
The only explicit dependency on the dimension $d$ is in the latter. The admissible fluxes and sources, i.e.~non-zero ones, depend implicitly on this dimension. These three equations will be key in the following derivations.

\subsection{No-go theorems in arbitrary dimension $d$}\label{sec:10dnogo}

We combine in the following the three equations \eqref{eq:dilatoneom}, \eqref{eq:10deinstein} and \eqref{eq:dRicciScalar} to reach, upon assumptions, an obstruction for having $\R_d > 0$. We get this way a no-go theorem against a solution with a $d$-dimensional de Sitter spacetime. We mostly generalize derivations of \cite{Andriot:2016xvq, Andriot:2017jhf} in $d=4$ to arbitrary dimension $d\geq 3$.

\subsubsection{Extension of Maldacena-Nuñez} \label{sec:MN}

Thanks to the dilaton equation of motion \eqref{eq:dilatoneom}, we eliminate the contribution of the Ricci scalar $\R_{10}$ in the 10d Einstein trace \eqref{eq:10deinstein} towards
\beq
\label{eq:dil10dcomb}
g_s \sum_p \frac{p-3}{p+1} T_{10}^{(p)} + 2|H|^2 - 2|H_7|^2 - g_s^2 \sum_{q=0}^7 (5-q) |F_q|^2=0 \ ,
\eeq
and use this result in \eqref{eq:dRicciScalar} to eliminate the $H$-flux. We obtain
\beq
\label{eq:dRicciScalar2}
\R_d= \frac{d}{4} \left( g_s \sum_p \frac{T_{10}^{(p)}}{p+1} - g_s^2 \sum_{q=0}^7 |F_q|^2 - 2 |H_7|^2 \right) \ .
\eeq
This provides an extension of the Maldacena-Nuñez no-go theorem \cite{Maldacena:2000mw} to arbitrary dimension $d \geq 3$, and to sources of any (possibly multiple) dimensionalities. We formulate it here with the following requirement
\beq
\label{eq:MNnogo}
\boxed{\text{De Sitter solutions require $T_{10}^{(p)}>0$ for some $p$.}}
\eeq
This is typically achieved by including $O_p$, whose contribution should dominate that of $D_p$.

\subsubsection{No-go for $p=7,8,9$, or $p=4,5,6$ with $F_{6-p}=0$, or $p=2$ with $H=0$} \label{sec:NonVanFlux}

This no-go theorem was derived in 4d in \cite{Wrase:2010ew} and in 10d in \cite{Andriot:2016xvq}, for $d=4$. It considers sources of single dimensionality $p$ (possibly intersecting, i.e.~in different sets). So we restrict here to a single $p$, giving $T_{10}=T_{10}^{(p)}$, and we recall that $p\geq d-1$. We first combine \eqref{eq:dilatoneom} with \eqref{eq:10deinstein} to eliminate $T_{10}$, and get
\beq
\label{eq:E10w/oT10}
\left(-2 \R_{10}+ |H|^2 - |H_7|^2 \right) (p-3) + 2 (|H|^2 - |H_7|^2) - g_s^2 \sum_{q=0}^7 (5-q) |F_q|^2 = 0 \ .
\eeq
Similarly, we combine \eqref{eq:dilatoneom} with \eqref{eq:dRicciScalar2} to eliminate $T_{10}$, resulting in
 \beq
 \label{eq:dRicciScalar3}
\left( 2+\frac{4}{d} \right) \R_d = -2 \R_{10-d} -  g_s^2 \sum_{q=0}^7 |F_q|^2 + |H|^2 - 3|H_7|^2  \ .
 \eeq
We now multiply \eqref{eq:dRicciScalar3} by $(p-3)$, insert \eqref{eq:E10w/oT10} and get
\beq
\label{eq:singlesizecomb}
\frac{4(p-3)}{d}\R_d = g_s^2 \sum_{q=0}^7 (8-p-q)|F_q|^2 - 2 |H|^2  + 2(4-p) |H_7|^2 \ .
\eeq
This extends \cite[(3.3)]{Andriot:2016xvq} to arbitrary dimension $d \geq 3$. More explicitly in IIA/B, we obtain
\bea
\frac{4 (p-3)}{d} \R_d &= g_s^2 \left( (8-p) |F_0|^2 + (6-p) |F_2|^2 + (4-p) |F_4|^2 + (2-p) |F_6|^2 \right) - 2 |H|^2 + 2(4-p) |H_7|^2 \nn\\
\frac{4 (p-3)}{d} \R_d &= g_s^2 \left( (7-p) |F_1|^2 + (5-p) |F_3|^2 + (3-p) |F_5|^2 + (1-p) |F_7|^2 \right) - 2 |H|^2 + 2(4-p) |H_7|^2 \nn
\eea
We first conclude for $p = 7, 8$ or $9$
\beq
\label{eq:p789nogo}
\boxed{\text{There is no de Sitter solution for $p = 7, 8$ or $9$ in any $d \geq 3$.}}
\eeq
We turn to $p=4, 5$ or $6$ as follows
\begin{itemize}
\item $p=6$:  evaluating equation \eqref{eq:singlesizecomb}, we get (as in \cite{Andriot:2016xvq, Andriot:2010ju} for $d=4$)
\beq
\frac{12}{d} \R_d = g_s^2 \left( 2 |F_0|^2 - 2 |F_4|^2 - 4 |F_6|^2 \right)- 2 |H|^2 -4 |H_7|^2 \ .
\eeq
For de Sitter, one needs $F_0 \neq 0$.
\item $p=5$: evaluating equation \eqref{eq:singlesizecomb}, we get (as in \cite{Andriot:2016xvq} for $d=4$)
\beq
\frac{8}{d} \R_d = g_s^2 \left( 2 |F_1|^2 - 2 |F_5|^2 - 4 |F_7|^2 \right)- 2 |H|^2 -2 |H_7|^2 \ .
\eeq
For de Sitter, one needs $F_1 \neq 0$.
\item $p=4$: evaluating equation \eqref{eq:singlesizecomb}, we get (as in \cite{Andriot:2016xvq, Andriot:2015aza} for $d=4$)
\beq
\frac{4}{d} \R_d = g_s^2 \left( 4 |F_0|^2 + 2 |F_2|^2 - 2 |F_6|^2 \right)- 2 |H|^2 \ .
\eeq
For de Sitter, one needs a priori $F_0 \neq 0$ or $F_2 \neq 0$. As pointed-out in \cite{Andriot:2018ept}, when having only $p=4$ sources, the $F_0$ Bianchi identity imposes a constant $F_0$, and the $O_4$ projection then implies $F_0=0$. So we deduce the only requirement of $F_2 \neq 0$.
\end{itemize}
We conclude (for a single $p\geq d-1$)
\beq
\label{eq:p456nogo}
\boxed{\text{There is no de Sitter solution for $p = 4, 5$ or $6$ in any $d \geq 3$ with $F_{6-p}=0$.}}
\eeq
Nothing can be said for $p=3$, but a different no-go theorem can be seen for $p=2$ (which occurs only for $d=3$)
\begin{itemize}
\item $p=2$:  evaluating equation \eqref{eq:singlesizecomb}, we get
\beq
-\frac{4}{d} \R_d = g_s^2 \left( 6 |F_0|^2 + 4 |F_2|^2 + 2 |F_4|^2 \right) - 2 |H|^2 + 4 |H_7|^2 \ .
\eeq
For de Sitter, one needs $H \neq 0$, meaning
\end{itemize}
\beq
\label{eq:p2nogo}
\boxed{\text{There is no de Sitter solution for $p = 2$ in $d = 3$ with $H=0$.}}
\eeq
Up to our knowledge, this is a new result, which will play an important role in the next sections.

Let us finally extend the derivation of \eqref{eq:singlesizecomb} to multiple dimensionalities $p,p'$. Proceeding similarly, we obtain the expression
\beq
\label{eq:multsizecomb}
\frac{4(p-3)}{d}\R_d = g_s^2 \sum_{q=0}^7 (8-p-q)|F_q|^2 - 2 |H|^2  + 2(4-p) |H_7|^2 + g_s \sum_{p'\neq p} \frac{p-p'}{p'+1} T_{10}^{(p')} \ .
\eeq

\subsubsection{Positive or vanishing internal curvature $\R_{10-d}$} \label{sec:PosCurv}

This no-go theorem was derived in 4d in \cite{Wrase:2010ew} and in 10d in \cite{Andriot:2016xvq}, for $d=4$. It is mentioned for arbitrary $d$ in \cite{VanRiet:2011yc}. This no-go is again about sources of single dimensionality $p$, to which we restrict ourselves. We combine \eqref{eq:dRicciScalar3} and \eqref{eq:singlesizecomb} to eliminate the $|H|^2$ term
\beq
\label{eq:singlesizecomb2}
\frac{d+p-1}{d} \R_d = - \R_{10-d} + \frac{g_s^2}{4} \sum_{q=0}^7 (6-p-q) |F_q|^2 + \frac{1-p}{2} |H_7|^2 \ .
\eeq
More explicitly, we obtain in IIA and IIB
\bea
\frac{d+p-1}{d} \R_d &= -\R_{10-d} + \frac{g_s^2}{4} \left( (6-p) |F_0|^2 + (4-p) |F_2|^2 + (2-p) |F_4|^2 -p |F_6|^2 \right) + \frac{1-p}{2} |H_7|^2 \nn \\
\frac{d+p-1}{d}  \R_d &= - \R_{10-d} + \frac{g_s^2}{4} \left( (5-p) |F_1|^2 + (3-p) |F_3|^2 + (1-p) |F_5|^2 - (1+p) |F_7|^2 \right) + \frac{1-p}{2} |H_7|^2 \nn
\eea
We conclude (for a single $p\geq d-1$)
\beq
\label{eq:R10-dnogo}
\boxed{\text{There is no de Sitter solution for $p \geq 4$ in any $d \geq 3$ with $\R_{10-d} \geq 0$.}}
\eeq
For $p=4$, this result requires the above argumentation on the $F_0$ flux.

\subsubsection{No-go for $p=d-1$} \label{sec:d-1}

We follow and extend the derivation in \cite{Andriot:2016xvq} to arbitrary $d\geq3$. This no-go theorem is at first about sources of a single dimensionality $p$, with $p$ being the minimal one, namely $p=d-1$. In that case, all internal dimensions are transverse to the sources: ${\rm vol}_{\bot} = {\rm vol}_{10-d}$. The sourced Bianchi identity
\beq
\label{eq:BI}
\d F_{8-p} - H \wedge F_{6-p} = \varepsilon_p \frac{T_{10}}{p+1} \text{vol}_{\perp} \ ,\quad  \varepsilon_p= (-1)^{p+1} (-1)^{\left[\frac{9-p}{2}\right]} \ ,
\eeq
can be reformulated as in \cite{Andriot:2016xvq,Andriot:2017jhf} by projecting on this volume, towards the scalar expression
\beq
2 g_s \frac{T_{10}}{p+1} = - \left|*_{10-d} H + \varepsilon_p g_s F_{6-p} \right|^2 + |H|^2 + g_s^2 |F_{6-p}|^2 + 2 \varepsilon_p g_s \left(\d F_{8-p} \right) \label{BIscalar}
\eeq
where $\d F_{8-p} = \left(\d F_{8-p} \right) {\rm vol}_{10-d}$.

Before using \eqref{BIscalar}, we consider the combination of \eqref{eq:dil10dcomb} and $-(p+1)\frac{4}{d}$ \eqref{eq:dRicciScalar2} for a single $p$
\beq
-(p+1) \frac{4}{d} \R_d =  -4 g_s \frac{T_{10}}{p+1} + 2 |H|^2 + 2 p |H_7|^2 + g_s^2 \sum_{q=0}^7 (p+q-4) |F_q|^2 \ . \label{pd-1mul0}
\eeq
It can be rewritten as follows, for any $2\leq p \leq 9$, keeping in mind that $F_q=0$ for $q<0$ or $q>7$
\bea
\R_d = -\frac{d}{2(p+1)} \Big( -2 g_s \frac{T_{10}}{p+1} + |H|^2 + &\, p |H_7|^2 + g_s^2 (  - |F_{2-p}|^2 + |F_{6-p}|^2 + 2 |F_{8-p}|^2 \label{eq:riccinogo1}\\
&  + 3 |F_{10-p}|^2 + 4 |F_{12-p}|^2 + 5 |F_{14-p}|^2 + 6 |F_{16-p}|^2)\Big) \ .\nn
\eea
Combined with \eqref{BIscalar}, we obtain
\bea
\label{eq:riccinogo2}
\R_d = -\frac{d}{2(p+1)} \bigg( &-2 g_s \varepsilon_p \left(\d F_{8-p} \right) + \left|*_{10-d} H + \varepsilon_p g_s F_{6-p} \right|^2 + p |H_7|^2 \\&
+g_s^2\left( - |F_{2-p}|^2 + 2 |F_{8-p}|^2 + 3 |F_{10-p}|^2 + 4 |F_{12-p}|^2 + 5 |F_{14-p}|^2 + 6 |F_{16-p}|^2 \right) \bigg) \nn
\eea
We finally consider $\int_{\mmm} {\rm vol}_{10-d}\, \R_d = \R_d \int_{\mmm}  {\rm vol}_{10-d} > 0 $ for a de Sitter solution. Focusing on $p=d-1$, one has $F_{8-p}= F_{9-d}$ and
\beq
\label{eq:MaxFluxCond}
\int_{\mmm} \d F_{9-d} = \int_{\del \mmm = 0} F_{9-d} = 0 \ .
\eeq
The integral of the right-hand side of \eqref{eq:riccinogo2} is thus negative for $p>2$, i.e.~$d>3$, leading to a no-go theorem for de Sitter. For $p=2$, the term $|F_{2-p}|^2$ is however contributing with the opposite sign. We conclude, for a single $p$
\beq
\label{eq:d-1nogo}
\boxed{\text{There is no de Sitter solution for $p = d-1$ in any $d\geq 4$.}}
\eeq
In addition, with a single $p$
\beq
\label{eq:p2F0nogo}
\boxed{\text{There is no de Sitter solution for $p = 2$ in $d=3$ with $F_0=0$.}}
\eeq

We finally extend the analysis to sources of multiple dimensionalities $p,p'$. If we consider as above the combination of \eqref{eq:dil10dcomb} and $-(p+1)\frac{4}{d}$ \eqref{eq:dRicciScalar2} for a given $p$, we extend \eqref{pd-1mul0} to
\beq
\hspace{-0.2in} -(p+1) \frac{4}{d} \R_d =  -4 g_s \frac{T_{10}^{(p)}}{p+1} + g_s \sum_{p'\neq p} \frac{p'-p-4}{p'+1} T_{10}^{(p')}  + 2 |H|^2 + 2 p |H_7|^2 + g_s^2 \sum_{q=0}^7 (p+q-4) |F_q|^2 \ . \label{eq:p'd-1}
\eeq
We can proceed as before, using the BI for $p=d-1$, while having contributions of sources with $p'>p$. The extra term leads us to conclude, for multiple dimensionalities (with $p'>p$)
\beq
\label{eq:p'd-1nogo}
\hspace{-0.4in} \boxed{\text{There is no de Sitter solution for $p = d-1$ in any $d\geq 4$ with $(p'-p-4) T_{10}^{(p')} \geq 0$ $\forall p'$.}}
\eeq
This was mentioned for $d=4, p=3, p'=5,7$ in \cite[(6.18)]{Andriot:2017jhf}. This formulation of the no-go theorem could be relaxed to requiring  $\sum_{p'\neq p} \frac{p'-p-4}{p'+1} T_{10}^{(p')} \geq 0$; in practice it is typically sufficient to consider each term of the sum separately as in \eqref{eq:p'd-1nogo}.

\subsubsection{Heterotic at order $(\alpha')^0$} \label{sec:Het}

Heterotic string at order $(\alpha')^0$ is effectively the NSNS sector of type IIA/B. The $d$-dimensional Einstein trace \eqref{eq:dRicciScalar} then boils down to
\beq
 \R_d= - \frac{d}{8} \left( |H|^2 + 3|H_7|^2 \right) \ .
\eeq
Using \eqref{eq:dRicciScalar2} we eliminate the contribution of $H_7$
\beq
\R_d=-\frac{d}{2}|H|^2 \ . \label{het10d}
\eeq
We conclude
\beq
\label{eq:Hetnogo}
\boxed{\text{There is no de Sitter solution in any $d \geq 3$ in heterotic string at order $(\alpha')^0$.}}
\eeq

Let us make a side remark. We recall that $H_7=0$ for $d\geq 4$. Comparing the above expressions, we deduce that one can only get Minkowski solutions for $d\geq 4$ in heterotic string at order $(\alpha')^0$. We note however that $d=3$ seems to allow for an anti-de Sitter solution as well, with $H, H_7 \neq 0$. It would correspond to a Freund-Rubin solution (see e.g.~\cite{Aguilar-Gutierrez:2022kvk}), but interestingly, a pure NSNS one. Such a solution on $AdS_3 \times S^3 \times T^4$ is for instance mentioned in \cite{Hoare:2013pma}.

\section{Interlude: excluding or constraining (quasi-) de Sitter in each dimension $d$}\label{sec:dSineachd}

In Section \ref{sec:dnogo}, we will reproduce and extend the no-go theorems obtained in Section \ref{sec:10dnogo} to forbid quasi-de Sitter solutions, using a $d$-dimensional approach. Here in this section, we already make a concrete use of these no-go theorems, to constrain, if not exclude, such solutions in each dimension $d \geq 3$. To that end, we determine in each $d$ the actual flux and source content: because the latter can be scarce, assumptions of the no-go theorems sometimes get automatically satisfied, allowing us to conclude. We first constrain this way de Sitter solutions in Section \ref{sec:dSineachdfirst}. We then restrict the discussion to supersymmetry-preserving source configurations in Section \ref{sec:susy}. We summarize our findings in Section \ref{sec:dSineachdsum} and finally show, in Section \ref{sec:quasidS}, that all these results are also valid for {\sl quasi}-de Sitter solutions.

A pioneering paper on these matters is \cite{VanRiet:2011yc}, where constraints in $d\geq 5$ on {\sl metastable} de Sitter solutions {\sl with supersymmetry-preserving source configurations} are discussed. In this section, we reproduce the (existence) results obtained there, and go beyond them with a more general analysis. Most of our results for $d>7$, and some of them for $5 \leq d \leq 7$, were already obtained in \cite{VanRiet:2011yc}. However, with the extra no-go theorem \eqref{eq:d-1nogo} at our disposal and further arguments, we exclude de Sitter solutions in some cases of $5 \leq d \leq 7$, and not just minima as in \cite{VanRiet:2011yc}. We provide further comparison along the text.

\subsection{First existence constraints}\label{sec:dSineachdfirst}

\subsection*{$d=10$}

\begin{table}[H]
  \begin{center}
    \begin{tabular}{|c|c|c|}
    \hline
Theory & Fluxes & $O_p/D_p$ dimensionality $p$ \\
    \hhline{===}
IIA & $F_0$ & $\varnothing$ \\
IIB & $\varnothing$ & 9 \\
    \hline
    \end{tabular}
  \end{center}
\end{table}

The absence of source in IIA leads to no-go theorem \eqref{eq:MNnogo}, and the only $p=9$ sources in IIB lead to no-go theorem \eqref{eq:p789nogo}. We conclude
\beq
\boxed{\text{There is no de Sitter solution in $d=10$.}}
\eeq

\subsection*{$d=9$}

\begin{table}[H]
  \begin{center}
    \begin{tabular}{|c|c|c|}
    \hline
Theory & Fluxes & $O_p/D_p$ dimensionality $p$ \\
    \hhline{===}
IIA & $F_0$ & 8 \\
IIB & $F_1$ & 9 \\
    \hline
    \end{tabular}
  \end{center}
\end{table}

The presence of only $p=8$ sources in IIA and $p=9$ sources in IIB lead to no-go theorem \eqref{eq:p789nogo}. We conclude
\beq
\boxed{\text{There is no de Sitter solution in $d=9$.}}
\eeq

\subsection*{$d=8$}

\begin{table}[H]
  \begin{center}
    \begin{tabular}{|c|c|c|}
    \hline
Theory & Fluxes & $O_p/D_p$ dimensionality $p$ \\
    \hhline{===}
IIA & $F_0, F_2$ & 8 \\
IIB & $F_1$ & 7,9 \\
    \hline
    \end{tabular}
  \end{center}
\end{table}

The presence of only $p=8$ sources in IIA leads to the no-go theorem \eqref{eq:p789nogo}. Considering sources of only one dimensionality $p$ in IIB would lead to the same conclusion. However, with both sources (not considered in \cite{VanRiet:2011yc}), the analysis is more involved. With the sources and fluxes of IIB, we can combine \eqref{eq:dil10dcomb} and \eqref{eq:dRicciScalar2} towards
\beq
\R_{8}=  - g_s \frac{T_{10}^{(9)}}{10} \label{d=8simplif}
\eeq
implying the de Sitter requirement $T_{10}^{(9)} < 0$. This leads to the no-go theorem \eqref{eq:p'd-1nogo} with $p'=9, p=7$; see also \eqref{T109=0}. We conclude
\beq
\boxed{\text{There is no de Sitter solution in $d=8$.}}
\eeq

\subsection*{$d=7$}

\begin{table}[H]
  \begin{center}
    \begin{tabular}{|c|c|c|}
    \hline
Theory & Fluxes & $O_p/D_p$ dimensionality $p$ \\
    \hhline{===}
IIA & $F_0, F_2, H$ & 6,8 \\
IIB & $F_1, F_3, H$ & 7,9 \\
    \hline
    \end{tabular}
  \end{center}
\end{table}

Let us start with IIA. Having $p=8$ sources only leads to no-go theorem \eqref{eq:p789nogo} and $p=6$ only leads to no-go theorem \eqref{eq:d-1nogo}; this leaves the possibility of having both together. In that case, the combination of \eqref{eq:dil10dcomb} and \eqref{eq:dRicciScalar2} leads to
\beq
\R_7 = \frac{7}{6}\left( -|H|^2+ g_s^2 |F_0|^2 - g_s \frac{T_{10}^{(8)}}{9} \right) = \frac{7}{10}\left( -|H|^2 - g_s^2 |F_2|^2 + g_s \frac{T_{10}^{(6)}}{7} \right) \ ,\label{R7}
\eeq
from which we deduce $T_{10}^{(6)}>0$ for de Sitter; this can also be read from \eqref{eq:multsizecomb} with $p=8$. No-go theorem \eqref{eq:p'd-1nogo} requires $T_{10}^{(8)} >0$ to have a de Sitter solution. We then deduce from \eqref{R7} the further requirement of $F_0 \neq 0$.

$T_{10}^{(6)}>0$ and $T_{10}^{(8)} >0$ mean that there are $O_6$ and $O_8$. The latter admits one transverse internal dimension. Along that direction, an $O_8$ is located at the fixed points of the orientifold involution. With respect to those fixed points, $F_0$ must be an odd function, because of this $O_8$ involution. The transverse space of the $O_6$ is the whole internal space. The $O_6$ involution applied to this transverse space requires $F_0$ to be an even function. The requirements of both $O_p$-planes can be found compatible with certain functions $F_0$, so we simply conclude
\beq
\hspace{-0.15in} \boxed{\text{There is no de Sitter solution in $d=7$ in type IIA, unless one has $O_6$, $O_8$ and $F_0\neq0$.}} \label{nodSd=7IIA}
\eeq
Note for instance that a constant $F_0$ would not satisfy the above constraints. A similar situation was discussed in \cite[Sec. 3.3.1]{Andriot:2022way} with $O_4$ and $O_6$, in $d=4$. There however, $F_0=0$ was possible, and de Sitter solutions were indeed obtained in the class $m_{46}^+$.\\

We turn to type IIB. Considering sources of only one dimensionality $p$ in IIB would lead to the no-go theorem \eqref{eq:p789nogo}. With both $p=7,9$ sources, the analysis is more involved. The combination of \eqref{eq:dil10dcomb} and \eqref{eq:dRicciScalar2} leads to
\beq
\R_7=-\frac{7}{8}\left(g_s \frac{T_{10}^{(9)}}{10} + g_s^2 |F_3|^2 + |H|^2 \right)=\frac{7}{12}\left( g_s \frac{T_{10}^{(7)}}{8} - g_s^2 \left( |F_1|^2 + 2 |F_3|^2\right) - |H|^2\right) \ ,
\eeq
from which we deduce $T_{10}^{(7)}>0$ and $T_{10}^{(9)}<0$ for a de Sitter solution; we reach the same result with \eqref{eq:multsizecomb}. This implies the presence of $O_7$, but not necessarily that of $O_9$; still it requires $D_9$. Note already that \eqref{T109=0} contradicts the requirement $T_{10}^{(9)}<0$; let us give a further argument. Since the Bianchi identity of $F_1$ is simply given by $\d F_1$, proportional to $T_{10}^{(7)}$, then we must have $F_1 \neq 0$. The $O_9$ projection has the particularity to always require $F_1=0$, since an $O_9$ fills the whole space while having the involution $\sigma_{O_9}(F_1)=-F_1$. We can then only have $D_9$ here. As discussed around \eqref{T109=0}, having $D_9$ alone is however not permitted. We conclude
\beq
\hspace{-0.2in} \boxed{\text{There is no de Sitter solution in $d=7$ in type IIB.}} \label{nodSd=7IIB}
\eeq
This excludes completely de Sitter solutions in $d=7$.

\subsection*{$d=6$}

\begin{table}[H]
  \begin{center}
    \begin{tabular}{|c|c|c|}
    \hline
Theory & Fluxes & $O_p/D_p$ dimensionality $p$ \\
    \hhline{===}
IIA & $F_0, F_2, F_4, H$ & 6,8 \\
IIB & $F_1, F_3, H$ & 5,7,9 \\
    \hline
    \end{tabular}
  \end{center}
\end{table}

Let us start with type IIA. Having $p=8$ sources alone leads to no-go theorem \eqref{eq:p789nogo}. For de Sitter, one must then have $p=6$ sources. The combination of \eqref{eq:dil10dcomb} and \eqref{eq:dRicciScalar2} leads to
\beq
\R_6 = \frac{3}{5}\left( g_s \frac{T_{10}^{(6)}}{7} -|H|^2- g_s^2 \left( |F_2|^2 + 2 |F_4|^2 \right) \right) = - g_s \frac{T_{10}^{(8)}}{9} -|H|^2  + g_s^2 \left( |F_0|^2 - |F_4|^2 \right) \ ,\label{R6IIA}
\eeq
from which we deduce $T_{10}^{(6)}>0$ and $g_s^2 |F_0|^2 - g_s \frac{T_{10}^{(8)}}{9} > 0$ for de Sitter; in particular one needs $O_6$. The second inequality indicates the need of $F_0 \neq0$: indeed, the $F_0$ Bianchi identity gives $\d F_0$ proportional to $T_{10}^{(8)}$, so $F_0$ cannot vanish while satisfying this inequality. Let us now consider the possibility of having $O_8$. In case the $O_8$ and $O_6$ are overlapping, then they share a common transverse direction along which they are localized. This situation is similar to the one in $d=7$, and there is no general obstruction. If they are not overlapping, the $O_8$ is then localized at a fixed point on the $O_6$, and its involution forces $F_0$ to be an odd function along that $O_6$ direction. This is not in contradiction with the $O_6$ involution constraint $\sigma_{O_6}(F_0)=F_0$, since the involution $\sigma_{O_6}$ has a trivial action along the $O_6$. Therefore, for non-overlapping $O_6$ and $O_8$, we do not encounter an issue either. Keeping in mind these different source configurations for Section \ref{sec:susy}, we conclude for now
\beq
\hspace{-0.2in} \boxed{\text{There is no de Sitter solution in $d=6$ in type IIA, unless one has $O_6$ and $F_0 \neq 0$.}}\label{nodSd=6IIA}
\eeq
\\

We turn to type IIB. In presence of sources of only one $p$, one cannot get de Sitter solutions: only $p=7$ or $9$ sources would lead to no-go \eqref{eq:p789nogo}, while $p=5$ leads to no-go \eqref{eq:d-1nogo}. We then consider sources of multiple dimensionalities. No-go theorem \eqref{eq:p'd-1nogo} forbids to have only $p=5$ with $p=9$, and requires to have $T_{10}^{(7)} > 0$ if there are $p=5$ sources. If there are no $p=5$ sources, the related equation \eqref{eq:p'd-1} still requires $T_{10}^{(7)} > 0$, as will be confirmed by equations below. So $O_7$ are always required. Having $T_{10}^{(7)} > 0$ requires in addition $F_1 \neq 0$ via its Bianchi identity, from which we deduce, as for $d=7$, the needed absence of $O_9$. As discussed already, having $D_9$ alone is not permitted, so overall, $p=9$ sources cannot be present here. The combination of \eqref{eq:dil10dcomb} and \eqref{eq:dRicciScalar2} leads to
\bea
\R_6 & = \frac{1}{2}\left( g_s \frac{T_{10}^{(7)}}{8}+ 2  g_s \frac{T_{10}^{(5)}}{6} -|H|^2- g_s^2 \left( |F_1|^2 + 2 |F_3|^2 \right) \right) \label{R6IIB} \\
& = \frac{3}{4}\left( -g_s \frac{T_{10}^{(9)}}{10}+  g_s \frac{T_{10}^{(5)}}{6} -|H|^2- g_s^2  |F_3|^2 \right)  \\
& = \frac{3}{2}\left( -2 g_s \frac{T_{10}^{(9)}}{10} -  g_s \frac{T_{10}^{(7)}}{8} -|H|^2 + g_s^2 |F_1|^2 \right)
\eea
where we should set $T_{10}^{(9)}=0$. We read from the above various conditions, in particular the need of $T_{10}^{(5)}>0$, therefore of $O_5$. We conclude
\beq
\hspace{-0.2in} \boxed{\text{There is no de Sitter solution in $d=6$ in type IIB, unless one has $O_7$, $F_1 \neq 0$ and $O_5$.}}\label{nodSd=6IIB}
\eeq

\subsection*{$d=5$}

\begin{table}[H]
  \begin{center}
    \begin{tabular}{|c|c|c|}
    \hline
Theory & Fluxes & $O_p/D_p$ dimensionality $p$ \\
    \hhline{===}
IIA & $F_0, F_2, F_4, H$ & 4,6,8 \\
IIB & $F_1, F_3, F_5, H$ & 5,7,9 \\
    \hline
    \end{tabular}
  \end{center}
\end{table}

In type IIA, having only $p=8$ sources leads to no-go \eqref{eq:p789nogo}, and only $p=4$ leads to no-go \eqref{eq:d-1nogo}. Having only $p=6$ sources seems possible, in which case one needs $O_6$ by \eqref{eq:MNnogo}. Turning to multiple dimensionalities, let us consider the no-go \eqref{eq:p'd-1nogo}: having $D_4/O_4$, with or without $D_8/O_8$, one gets the de Sitter requirement of having $T_{10}^{(6)} > 0$, i.e.~having $O_6$. In absence of $D_4/O_4$, one can still use the related equation \eqref{eq:p'd-1} with $p=4$ to conclude on the same need of $O_6$ with $T_{10}^{(6)} > 0$, whether or not there are $D_8/O_8$. We conclude
\beq
\hspace{-0.2in} \boxed{\text{There is no de Sitter solution in $d=5$ in type IIA, unless one has $O_6$.}}\label{nodSd=5IIA}
\eeq
As above, a few more constraints can be derived, although nothing decisive. Let us still add, for future purposes, that the combination of \eqref{eq:dil10dcomb} and \eqref{eq:dRicciScalar2} leads in particular to
\beq
\R_5  = \frac{5}{6}\left( g_s \frac{T_{10}^{(4)}}{5} - g_s \frac{T_{10}^{(8)}}{9} -|H|^2 + g_s^2 \left( |F_0|^2 - |F_4|^2 - 2 |F_6|^2 \right) \right) \ . \label{R5IIA}
\eeq
We deduce that in absence of $p=4$ sources, one must have $g_s^2 |F_0|^2  - g_s \frac{T_{10}^{(8)}}{9} > 0$. We are then in the same situation as in $d=6$, and one can run a similar discussion on $O_8/D_8$, whether those have a common transverse direction with the $O_6$ or not.\\

In type IIB, having sources of single dimensionality only allows $p=5$ due to no-go \eqref{eq:p789nogo}, and the need of $O_5$ by no-go \eqref{eq:MNnogo}. Having multiple dimensionalities allows a priori for more options: let us consider again the combination of \eqref{eq:dil10dcomb} and \eqref{eq:dRicciScalar2}, leading to
\bea
\R_5 & = \frac{5}{12}\left( g_s \frac{T_{10}^{(7)}}{8}+ 2  g_s \frac{T_{10}^{(5)}}{6} -|H|^2- g_s^2 \left( |F_1|^2 + 2 |F_3|^2 + 3 |F_5|^2  \right) \right) \label{R6IIB} \\
& = \frac{5}{8}\left( -g_s \frac{T_{10}^{(9)}}{10}+  g_s \frac{T_{10}^{(5)}}{6} -|H|^2- g_s^2 ( |F_3|^2 + 2 |F_5|^2  )\right)  \\
& = \frac{5}{4}\left( -2 g_s \frac{T_{10}^{(9)}}{10} -  g_s \frac{T_{10}^{(7)}}{8} -|H|^2 + g_s^2 (|F_1|^2 - |F_5|^2) \right)
\eea
where from \eqref{T109=0} we should set $T_{10}^{(9)}=0$. From this we first deduce the requirement $T_{10}^{(5)} >0$ for de Sitter, so the need of $O_5$ in the case of multiple dimensionalities as well. We also read the requirement $  g_s^2 |F_1|^2 -  g_s \frac{T_{10}^{(7)}}{8} > 0$. If $F_1=0$, this inequality cannot be satisfied because the $F_1$ Bianchi identity then gives $T_{10}^{(7)}=0$. We must then have $F_1 \neq 0$, which, as argued previously, forbids to have $O_9$, and thus $D_9$. In other words, de Sitter forbids $p=9$ sources and we conclude
\beq
\hspace{-0.2in} \boxed{\text{There is no de Sitter solution in $d=5$ in type IIB, unless one has $O_5$ and $F_1 \neq 0$.}}\label{nodSd=5IIB}
\eeq

\subsection*{$d=4$}

\begin{table}[H]
  \begin{center}
    \begin{tabular}{|c|c|c|}
    \hline
Theory & Fluxes & $O_p/D_p$ dimensionality $p$ \\
    \hhline{===}
IIA & $F_0, F_2, F_4, F_6, H$ & 4,6,8 \\
IIB & $F_1, F_3, F_5, H$ & 3,5,7,9 \\
    \hline
    \end{tabular}
  \end{center}
\end{table}

For completeness, we list $d=4$, which is by far the most studied case. Constraints for single dimensionalities have been discussed e.g.~in \cite{Andriot:2019wrs, Andriot:2018ept} and references therein, while some for multiple dimensionalities have appeared in \cite[Sec. 6]{Andriot:2017jhf} and more extensively in \cite{Andriot:2022way}. Single dimensionalities only allow $p=4,5$ or $6$; it was further argued that $p=4$ is very unlikely \cite{Andriot:2022way}. Multiple dimensionalities offer many options \cite{Andriot:2022way}.

\subsection*{$d=3$}

\begin{table}[H]
  \begin{center}
    \begin{tabular}{|c|c|c|}
    \hline
Theory & Fluxes & $O_p/D_p$ dimensionality $p$ \\
    \hhline{===}
IIA & $F_0, F_2, F_4, F_6, H, H_7$ & 2,4,6,8 \\
IIB & $F_1, F_3, F_5, F_7, H, H_7$ & 3,5,7,9 \\
    \hline
    \end{tabular}
  \end{center}
\end{table}

Many options seem again possible for de Sitter and we refrain from listing them all. Interestingly though, a few peculiarities appear in $d=3$: some no-go theorems do not work as in higher dimensions. In type IIA with only $p=2$ sources, one gets different constraints, as indicated in \eqref{eq:p2nogo} and \eqref{eq:p2F0nogo}, which will play an important role in the next sections.

Classical de Sitter solutions in $d=3$ have been discussed in \cite{Farakos:2020idt} and \cite{Emelin:2021gzx}. The former goes however slightly beyond our framework since anti-$D_p$ are considered in IIA; some features there remain interesting. The discussion in \cite{Emelin:2021gzx} is done in IIB and, interestingly, involves $F_7$-flux, without however finding de Sitter solutions.

\subsection{Restricting to supersymmetric source configurations}\label{sec:susy}

There are several motivations to require that the $O_p/D_p$ source configuration does not break supersymmetry: first, it leaves a chance to get a supersymmetric effective theory in $d$ dimensions, which could be phenomenologically interesting, and second, it avoids possible instabilities that would occur among the $O_p/D_p$. This requirement has also been made in the analysis of \cite{VanRiet:2011yc}; we reproduce and extend here the results obtained there.

Possible supersymmetric source configurations for a de Sitter solution in $d=4$ were discussed in \cite[Sec. 2.4.2]{Andriot:2022way}, and we consider here higher dimensions $d$. In the presence of 2 intersecting sets of orthogonal $O_p/D_p$, the rule is well-known: the total number of Neumann-Dirichlet boundary conditions should be a multiple of 4. We denote this number as ${\cal N}_{ND}$. This condition amounts to evaluate the total number of directions wrapped by one set but not by the other, and require it to be a multiple of 4. If this holds, then supersymmetry is only broken by a quarter; it is otherwise fully broken (sources are not mutually BPS), which is what we want to avoid in this section. Note that having anti-$D_p$-branes together with $O_p$ would also lead to such a breaking, which could be another argument not to include them.

Instead of systematically determining the configurations that would preserve supersymmetry for every dimension $d$, let us focus only on those for which a de Sitter solution is possible, as determined in Section \ref{sec:dSineachdfirst}. We thus start in $d=7$.

\subsection*{$d=7$}

The only possible configuration identified there was in type IIA with $O_6$ and $O_8$ \eqref{nodSd=7IIA}. Given an $O_6$ is fully space filling in $d=7$, it is completely along an $O_8$, so one gets ${\cal N}_{ND}=2$ and supersymmetry cannot be preserved. We conclude by phrasing this result as follows
\beq
\boxed{\text{There is no de Sitter solution in a $d=7$ susy theory.}}
\eeq

\subsection*{$d=6$}

Let us start with type IIA, with the requirement of $O_6$ \eqref{nodSd=6IIA}. Since an $O_6$ wraps only one internal dimension, there cannot be intersecting $O_6/D_6$ that preserve supersymmetry: those would lead to ${\cal N}_{ND}=2$. Turning to $O_8/D_8$, the only possibility to preserve supersymmetry is to place them as in Table \ref{tab:d=6IIAsusy}. Note that this is the case discussed above \eqref{nodSd=6IIA} where $O_8$ and $O_6$ are not overlapping.
\begin{table}[H]
  \begin{center}
    \begin{tabular}{|c||c||c|c|c|c|}
    \hline
Sources & $d$ spacetime dimensions & 1 & 2 & 3 & 4 \\
    \hhline{=::=::====}
  $O_6$, ($D_6$) & $\otimes$ & $\otimes$ & & & \\
    \hhline{-||-||----}
  ($O_8, D_8$) & $\otimes$ & & $\otimes$ & $\otimes$ & $\otimes$ \\
    \hline
    \end{tabular}
    \caption{Supersymmetry-preserving source configuration allowing for possible de Sitter solutions in type IIA in $d=6$}\label{tab:d=6IIAsusy}
  \end{center}
\end{table}

We turn to type IIB, with the requirement of $O_7$ and $O_5$ \eqref{nodSd=6IIB}. $O_5$ (and possibly $D_5$) are space-filling, so $O_7$ and $O_5/D_5$ necessarily lead here to ${\cal N}_{ND}=2$, and thus do not preserve supersymmetry.

We conclude by phrasing these results as follows
\beq
\boxed{\text{There is no de Sitter solution in a $d=6$ susy theory, unless from $O_p/D_p$ as in Table \ref{tab:d=6IIAsusy}.}}
\eeq
Let us nevertheless express doubts on the ability of the IIA configuration in Table \ref{tab:d=6IIAsusy} to provide a de Sitter solution. First, a T-duality along one of its directions would lead to $O_5, O_9, D_9$, or $O_7,D_7$, neither of which allows for de Sitter according to \eqref{nodSd=6IIB}. In addition, considering only $O_6$ then amounts to have a single set of parallel sources, a situation conjectured in $d=4$ to not allow for any de Sitter solution \cite{Andriot:2019wrs}. We nevertheless do not yet have no-go theorems to exclude this configuration. Let us add that searches for de Sitter solutions in $d=6$ were performed in \cite{Dibitetto:2019odu} for source configurations preserving $\frac{1}{2}$ supersymmetries, i.e.~one set of sources. No de Sitter solution was found in our setting, in particular with $O_6/D_6$. De Sitter solutions are nevertheless mentioned when having Kaluza--Klein monopoles $KKO_5/KK_5$ in type IIA; for those the compactness of the manifold is however not ensured.

\subsection*{$d=5$}

Let us start with type IIA, with the requirement of $O_6$, and possible $p=4,8$ sources \eqref{nodSd=5IIA}. $O_4/D_4$ sources cannot preserve supersymmetry together with $O_6$ since they are space-filling. We fall then in the case discussed below \eqref{R5IIA}, which may or may not include $O_8$. In addition, having two intersecting sets of $O_8/D_8$ would lead to ${\cal N}_{ND}=2$. The only possible supersymmetry preserving source configurations for de Sitter are thus those of Table \ref{tab:d=5IIAsusy}.
\begin{table}[H]
  \begin{center}
    \begin{tabular}{|c||c||c|c|c|c|c|}
    \hline
Sources & $d$ spacetime dimensions & 1 & 2 & 3 & 4 & 5 \\
    \hhline{=::=::=====}
  $O_6$, ($D_6$) & $\otimes$ & $\otimes$ & $\otimes$ & & & \\
    \hhline{-||-||-----}
  ($O_6, D_6$) & $\otimes$ & & & $\otimes$ & $\otimes$ &  \\
    \hline
  \multicolumn{7}{c}{}\\
    \hline
  $O_6$, ($D_6$) & $\otimes$ & $\otimes$ & $\otimes$ & & & \\
    \hhline{-||-||-----}
  ($O_8, D_8$) & $\otimes$ & & $\otimes$ & $\otimes$ & $\otimes$ & $\otimes$ \\
    \hline
    \end{tabular}
    \caption{Supersymmetry-preserving source configurations allowing for possible de Sitter solutions in type IIA in $d=5$}\label{tab:d=5IIAsusy}
  \end{center}
\end{table}

We turn to type IIB, for which $p=5,7$ sources are a priori possible, with the systematic need of $O_5$ \eqref{nodSd=5IIB}. For single dimensionality, we concluded that only $p=5$ could give de Sitter solutions. Since $O_5/D_5$ wrap only one internal dimension, supersymmetry preservation allows for one set at most. Turning to multiple dimensionalities with $p=5,7$, we find some possibility to preserve supersymmetry. Overall, the supersymmetry preserving source configurations are given in Table \ref{tab:d=5IIBsusy}. It seems that the combination of $p=5$ and $p=7$ sources has been missed in \cite{VanRiet:2011yc}.
\begin{table}[H]
  \begin{center}
    \begin{tabular}{|c||c||c|c|c|c|c|}
    \hline
Sources & $d$ spacetime dimensions & 1 & 2 & 3 & 4 & 5 \\
    \hhline{=::=::=====}
  $O_5$, ($D_5$) & $\otimes$ & $\otimes$ & & & & \\
    \hhline{-||-||-----}
  ($O_7, D_7$) & $\otimes$ & & $\otimes$ & $\otimes$ & $\otimes$ & \\
    \hline
    \end{tabular}
    \caption{Supersymmetry-preserving source configuration allowing for possible de Sitter solutions in type IIB in $d=5$}\label{tab:d=5IIBsusy}
  \end{center}
\end{table}

We conclude
\beq
\boxed{\text{There is no de Sitter solution in a $d=5$ susy theory, unless from $O_p/D_p$ as in Table \ref{tab:d=5IIAsusy} and \ref{tab:d=5IIBsusy}.}}
\eeq
As for $d=6$, we nevertheless have doubts on the ability to get de Sitter solutions with these source configurations, even though we do not have decisive no-go theorems. Again, T-dualities on the source configurations lead to others for which de Sitter is not possible, while configurations with a single set of $O_5$ or $O_6$ also disfavour de Sitter.

\subsection{Summary and perspectives}\label{sec:dSineachdsum}

For each dimension $d$, we have determined the field and source content in type IIA/B compactifications. The scarcity of this content, especially for the highest $d$, sometimes implied that the assumptions of no-go theorems of Section \ref{sec:10dnogo} were automatically satisfied. This allowed to concretely exclude, or constrain, the existence of a solution with a $d$-dimensional de Sitter spacetime. More precisely, de Sitter solutions are forbidden for  $8 \leq d  \leq 10$ and in type IIB in $d=7$. For $d=7,6,5$, constraints were found in \eqref{nodSd=7IIA}, then \eqref{nodSd=6IIA}, \eqref{nodSd=6IIB} and \eqref{nodSd=5IIA}, \eqref{nodSd=5IIB}. The cases of $d=4,3$ were also discussed and referenced in Section \ref{sec:dSineachdfirst}. When restricting in addition to supersymmetry-preserving source configurations in Section \ref{sec:susy}, de Sitter got fully excluded in $d=7$ and in type IIB in $d=6$. Only few source configurations could then still allow for de Sitter: those of Table \ref{tab:d=6IIAsusy} in $d=6$, and Table \ref{tab:d=5IIAsusy} and  \ref{tab:d=5IIBsusy} in $d=5$, while $d=4$ was treated in \cite[Sec. 2.4.2]{Andriot:2022way}. These results on the (non)-existence of de Sitter in all dimensions $d \geq 3$ reproduce and extend those of the pioneering paper \cite{VanRiet:2011yc}. They will be extended to quasi-de Sitter solutions in Section \ref{sec:quasidS}.

Restricting to supersymmetric source configurations is a reasonable assumption for phenomenology, since it can avoid some instabilities, and possibly provides a supersymmetric $d$-dimensional effective theory. Doing so, we are only left with few source configurations in $d=6,5$. Interestingly, we note that those contain at most 2 sets of intersecting $O_p/D_p$ sources: see Table \ref{tab:d=6IIAsusy}, \ref{tab:d=5IIAsusy}, \ref{tab:d=5IIBsusy}. This is expected due to the number of supercharges: for $d\geq 5$, the supersymmetry algebra does not allow for only 4 supercharges.\footnote{We thank N.~Cribiori for related discussion.} Having 3 intersecting sets of sources would break at least supersymmetry to $\frac{1}{8}$ of the original amount, here 32 supercharges in type IIA/B. Preserving supersymmetry in $d\geq 5$ then allows at most for 2 intersecting sets, as observed here.

This observation is crucial in regards of Conjecture 4 of \cite{Andriot:2022way}: it states that de Sitter solutions cannot be found with 2 sets of intersecting sources. Conjecture 1 of \cite{Andriot:2019wrs} equally forbids de Sitter solutions with 1 set, i.e.~parallel $O_p/D_p$, meaning that overall, de Sitter solutions need at least 3 sets, giving a maximum of 4 supercharges \cite{Andriot:2022way}. Since the supersymmetric source configurations remaining in Section \ref{sec:susy} have at most 2 sets, proving Conjecture 1 and 4 would exclude de Sitter solutions in $d \geq 5$. It would leave $d=4$ as the highest possible dimension for de Sitter, an exciting possibility.

\subsection{Extension to quasi-de Sitter}\label{sec:quasidS}

We show in the following that all results obtained in Section \ref{sec:dSineachdfirst} and \ref{sec:susy}, constraining or excluding de Sitter solutions in each dimension $d$, are straightforwardly extended to quasi-de Sitter solutions, using a $d$-dimensional effective theory discussed in Section \ref{sec:ddim} with scalar potential $V$ given in \eqref{pot}. In that language, a de Sitter solution is a critical point $\del_{\varphi} V =0$ with $V>0$, and a quasi-de Sitter one is a setting with $V>0$ and a slope $|\del_{\varphi} V |$ (here for single field) not too large, as will be specified below. This extension is important for cosmology.\\

Anticipating on \eqref{pot}, one can show that
\bea
& -\frac{2}{M_p^2} \left( \frac{2d}{d-2}\, V + \frac{2}{d-2} \tau \del_\tau V \right) \label{eq:dild} \\
& = 2 \tau^{-2} \rho^{-1} \R_{10-d} + \, \tau^{-\frac{d+2}{2}}\, \sum_p \rho^{\frac{2p-8-d}{4}}\, g_s \frac{T_{10}^{(p)}}{p+1} - \tau^{-2} \rho^{-3}\, |H|^2 + \tau^{2(1-d)} \rho^{3-d} |H_7|^2  \nn
\eea
reproduces (on-shell, at the critical point where $\rho=\tau=1$) the dilaton e.o.m.~\eqref{eq:dilatoneom}, using the relation between ${\cal R}_d$ and $V$ \eqref{RdV}. Similarly, the following equation
\bea
& -\frac{2}{M_p^2} \left( \frac{4d}{d-2}\, V + \frac{10-d}{2(d-2)} \tau \del_\tau V + \rho \del_{\rho} V \right) \label{eq:10deinsteind} \\
& = 4 \tau^{-2} \rho^{-1} \R_{10-d} + \frac{1}{2}\, \tau^{-\frac{d+2}{2}}\, \sum_p \rho^{\frac{2p-8-d}{4}}\, g_s T_{10}^{(p)} - \tau^{-2} \rho^{-3}\, |H|^2 + \tau^{2(1-d)} \rho^{3-d} |H_7|^2 \nn\\
& - \frac{1}{2} \tau^{-d}\, g_s^2 \sum_{q=0}^{10-d} \rho^{\frac{10-d-2q}{2}}\, (5-q) |F_q|^2   \nn
\eea
reproduces (on-shell) the 10d Einstein trace \eqref{eq:10deinstein}, and
\bea
& \frac{2}{M_p^2} \left( \frac{d}{d-2}\, V + \frac{(10-d)d}{16(d-2)} \tau \del_\tau V + \frac{d}{8} \rho \del_{\rho} V \right) \label{eq:dRicciScalard} \\
& = \frac{d}{16}\Big(\, \tau^{-\frac{d+2}{2}}\, g_s\, \sum_p \rho^{\frac{2p-8-d}{4}}\, \frac{7-p}{p+1} T_{10}^{(p)} - 2 \tau^{-2} \rho^{-3}\, |H|^2 -6 \tau^{2(1-d)} \rho^{3-d} |H_7|^2 \nn\\
& + \tau^{-d}\, g_s^2 \sum_{q=0}^{10-d} \rho^{\frac{10-d-2q}{2}} (1-q)\, |F_q|^2  \Big) \nn
\eea
reproduces (on-shell) the $d$-dimensional Einstein trace \eqref{eq:dRicciScalar}. In other words, it can be seen as the same equations up to first order derivatives of $V$, and (positive) factors made of powers of fields $(\rho,\tau)$.

One can verify that all no-go theorems above and further constraints on the existence of de Sitter solutions in various dimensions $d$ have involved\\
- linear combinations of \eqref{eq:dilatoneom}, \eqref{eq:10deinstein} and \eqref{eq:dRicciScalar}\\
- Bianchi identities of $F_0, F_1, F_{9-d}$\\
- orientifold projections\\
- in Section \ref{sec:susy}, rules for supersymmetry preserving source configurations.\\
In addition, the impossibility of getting de Sitter was always reached as ${\cal R}_d \leq 0$ by reasoning on signs of {\sl individual} terms, such as each $T_{10}^{(p)}$, and not linear combinations thereof.

It is then clear that the entire same reasonings can be pursued using \eqref{eq:dild}, \eqref{eq:10deinsteind} and \eqref{eq:dRicciScalard}, in place of \eqref{eq:dilatoneom}, \eqref{eq:10deinstein} and \eqref{eq:dRicciScalar}. Indeed, one can consider the same linear combinations. The Bianchi identities still hold in a $d$-dimensional effective theory, and can be decorated by the appropriate scalar field factors if necessary, as in \eqref{BIscalar2}. The orientifold projections and the supersymmetry rules also still hold. Last but not least, it is crucial that the reasonings are based on the signs of individual terms: the scalar factors then do not alter the results. If on the contrary they were based on sums of terms which get different scalar factors, the reasonings and conclusions could have been altered. This last situation was referred to as ``field dependent condition'' in \cite{Andriot:2020lea}, and is precisely mentioned to be avoided in Section \ref{sec:dnogo} in the no-go theorems considered. In short, all previous reasonings excluding de Sitter in $d$ dimensions can be reproduced in a $d$-dimensional theory with scalar potential as explained above, the only difference being the first order derivatives of the potential. Instead of ${\cal R}_d \leq 0$, one then reaches, upon assumptions, conclusions of the form
\beq
\label{eq:nogolincomb0}
a\, V + \sum_{i} b_i\, \varphi^i \del_{\varphi^i} V  \leq 0 \ ,
\eeq
with $a>0$ and typically $\exists\, b_i \neq 0$. The precise derivation of the no-go theorems in $d$ dimensions in Section \ref{sec:dnogo} will be an illustration.

Interestingly, the inequality \eqref{eq:nogolincomb0} holds also off-shell (away from a de Sitter critical point where $\varphi^i=1$). Physically, it has the crucial consequence that not only de Sitter solutions are excluded in all dimensions $d$ and situations where \eqref{eq:nogolincomb0} is valid, but also quasi-de Sitter ones. The latter refers to having $V >0$ and a slope $|\nabla V|$ not too large. Indeed, as explained in e.g.~\cite[(1.5)]{Andriot:2021rdy} and reminded here in \eqref{eq:NoGoEq}, the inequality \eqref{eq:nogolincomb0} implies a lower bound on the slope: $|\nabla V| \geq c \, V$ ($M_p=1$). This is true, provided $\exists\, b_i \neq 0$ and the fields $\varphi^i$ can be brought to canonical ones, which is the case here; more details will be given in Section \ref{sec:dnogo}. In conclusion, all excluded or constrained de Sitter solutions in $d$ dimensions, as summarized in Section \ref{sec:dSineachdsum}, lead to the same exclusion or constraints on quasi-de Sitter solutions. For the latter, the bound is given by the number $c$, typically of order $1$. This conclusion is relevant for cosmology.

\section{$d$-dimensional derivations}\label{sec:ddim}

We perform in Section \ref{sec:dimred} a dimensional reduction from a $D$-dimensional spacetime towards a $d$-dimensional one, compute the kinetic terms of appearing scalar fields, as well as their scalar potential for $D=10$ and type IIA/B supergravity. We then use this $d$-dimensional effective theory in Section \ref{sec:dnogo} to reproduce the no-go theorems of Section \ref{sec:10dnogo}, that were obtained via 10d equations, and we consider two other no-go theorems. For all of those, we also compute the ($d$-dependent) $c$ value of the swampland de Sitter conjecture \cite{Obied:2018sgi}. These results will be summarized and discussed in Section \ref{sec:discussionC}.

\subsection{Dimensional reduction and effective action}\label{sec:dimred}

We consider two actions, in $D$ and $d$ dimensions, related as follows
\bea
{\cal S} & = \frac{1}{2 \kappa_{D}^2} \int \d^{D} x \sqrt{|g_{D}|} e^{-2\phi} \Big( \R_{D} + 4 (\del \phi)_{D}^2 \Big) \ ,\\
& = \int \d^d x \sqrt{|g_d|} \left(\frac{M_p^2}{2} \R_d - \frac{1}{2} g_{ij} \del_{\mu}\varphi^i \del^{\mu}\varphi^j - V \right) \ ,\label{Sdgen}\\
{\rm with}\ M_p^2 & = \frac{1}{\kappa_{D}^2} \int \d^{D-d} y \sqrt{|g_{D-d}|}\ g_s^{-2} \ . \label{Mp}
\eea
The $D$-dimensional action depends on the metric and a dilaton field $\phi$ (whose kinetic term is squared with that $D$-dimensional metric). $\kappa_{D}$ is a constant. The $d$-dimensional action is given in Einstein frame. Beyond the $d$-dimensional reduced Planck mass $M_p$ and metric $g_{\mu \nu}$, it depends on the field space metric $g_{ij}$ and a scalar potential $V$ for the scalar fields $\varphi^i$. In the following, we relate these two actions in more detail.

Before doing so, let us consider a solution of the $d$-dimensional theory, given by an extremum of the potential (without scalar kinetic energy). In that case, the trace of the $d$-dimensional Einstein equation gives the following relation
\beq
\frac{d-2}{d} \R_d = \frac{2}{M_p^2} V \ . \label{RdV}
\eeq
A de Sitter solution then corresponds to a positive extremum of the potential.

A first set of $d$-dimensional scalar fields to be considered will be $(\rho, \tau, \sigma)$ (definitions and references will be given below). The ansatz for dimensional reduction is then as follows: we split the $D$-dimensional metric into
\beq
\d s^2_{D} = \tau^{-2}(x)\, g_{\mu \nu}(x)\, \d x^{\mu} \d x^{\nu} + \rho(x)\, g_{mn}(x,y)\, \d y^m \d y^n\ ,\label{metricd}
\eeq
with $\mu =0, \dots, d-1,\; m=d, \dots, D-1$. The scalar fields should be viewed as fluctuations around a background metric. With this ansatz, one gets the relation $\R_D = \tau^2 \R_d + \rho^{-1} \R_{D-d}+ \dots$, where the two Ricci scalars are given in terms of $g_{\mu \nu}$ and $g_{mn}$, with $d$- and $(D-d)$-dimensional derivatives respectively. In addition, we consider $e^{\phi} = g_s \, e^{\delta \phi (x)}$, with a constant $g_s$. Going to the $d$-dimensional Einstein frame then requires to pick
\beq
e^{\delta \phi} = \tau^{-\frac{d-2}{2}} \rho^{\frac{D-d}{4}} \ ,
\eeq
which in turn fixes $M_p$ as in \eqref{Mp}. This leads to the scalar potential
\beq
V= \frac{1}{2\kappa_{D}^2} \int \d^{D-d} y \sqrt{|g_{D-d}|}\ g_s^{-2}\, \times (-\tau^{-2})\times (\rho^{-1} \R_{D-d}) \ , \label{pot0}
\eeq
and we will complete the latter with more ingredients below. We highlight the prefactor $(-\tau^{-2})$ coming from prefactors in the $D$-dimensional action and the dimensional reduction. We now turn to the scalar kinetic terms.

\subsubsection{Kinetic terms for $(\rho, \tau, \sigma)$}\label{sec:kin}

To compute the kinetic terms for $(\rho, \tau, \sigma)$, we generalize the results of \cite[App. D]{Andriot:2020wpp} to arbitrary dimensions $D$ and $d$. A first contribution to the kinetic terms comes from the dilaton
\beq
4 (\del \phi)_{D}^2 = \tau^2 \left( \frac{(D-d)^2}{4} \rho^{-2} (\del \rho)^2 + \frac{(d-2)^2}{4} \tau^4 (\del \tau^{-2})^2 + \frac{(d-2)(D-d)}{2} \tau^2 \rho^{-1} \del_{\mu} \tau^{-2} \del^{\mu} \rho \right) \ ,\nn
\eeq
where the squares on the right-hand side are now made with $g_{\mu \nu}$. The other, less trivial contribution comes from $\R_D$ (with Levi-Civita connection). A first computation leads to
\bea
\R_{D} & = \tau^2 \R_d + \rho^{-1} \R_{D-d} - \nabla_{\mu} \left( (d-1) \tau^4 \del^{\mu} \tau^{-2} + \tau^2 \rho^{-1} g^{mn} \del^{\mu} (\rho g_{mn}) \right) \\
& - \frac{(d+2)(d-1)}{4} \tau^6 (\del \tau^{-2})^2 - \frac{d}{2} \tau^4 \del_{\mu} \tau^{-2} \del^{\mu} (\rho g_{mn}) \rho^{-1} g^{mn} \nn\\
& - \frac{\tau^2}{4} \left( \rho^{-1} g^{mn} \del (\rho g_{mn}) \right)^2 + \frac{\tau^2}{4} \del_{\mu} (\rho g_{mn}) \del^{\mu} (\rho^{-1} g^{mn}) \nn\ ,
\eea
where $g_{\mu\nu}$ is used to define covariant derivatives, lift indices and in the squares. The next step is to perform an integration by parts of the previous derivative $\nabla_{\mu}$, while we have to take into account the prefactor $\tau^{-2}$ allowing to reach the Einstein frame. This leads to
\bea
& \int \d^d x \sqrt{|g_d|}\, \tau^{-2}\, \R_{D} \label{int1}\\
= & \int \d^d x \sqrt{|g_d|} \, \Big( \R_d + \tau^{-2} \rho^{-1} \R_{D-d} - \frac{(d-1)(d-2)}{4} \tau^4 (\del \tau^{-2})^2 - \frac{d-2}{2} \tau^2 \del_{\mu} \tau^{-2} \del^{\mu} (\rho g_{mn}) \rho^{-1} g^{mn} \nn\\
& \phantom{\int \d^d x \sqrt{|g_d|} \, ( } - \frac{1}{4}  \left( \rho^{-1} g^{mn} \del (\rho g_{mn}) \right)^2 + \frac{1}{4} \del_{\mu} (\rho g_{mn}) \del^{\mu} (\rho^{-1} g^{mn}) \Big) \nn\ .
\eea
Finally, we recall that $\del_{\mu} \ln {\rm det} M = {\rm Tr} (M^{-1} \del_{\mu} M)$ for an invertible matrix $M$. We assume from now on that $g_{D-d}={\rm det}\, g_{mn}$ is independent of coordinates $x$; this will be further justified below. This implies $g^{mn} \del_{\mu} g_{mn} = 0$, leading to the following expression
\bea
& \int \d^d x \sqrt{|g_d|}\, \tau^{-2}\, \R_{D} \label{int2}\\
= & \int \d^d x \sqrt{|g_d|} \, \Big( \R_d + \tau^{-2} \rho^{-1} \R_{D-d} - \frac{(d-1)(d-2)}{4} \tau^4 (\del \tau^{-2})^2 - \frac{(d-2)(D-d)}{2} \tau^2 \rho^{-1} \del_{\mu} \tau^{-2} \del^{\mu} \rho \nn\\
& \phantom{\int \d^d x \sqrt{|g_d|} \, (} - \frac{(D-d)^2}{4} \rho^{-2} (\del \rho )^2 - \frac{D-d}{4} \rho^{-2} (\del \rho)^2 + \frac{1}{4} \del_{\mu} (g_{mn}) \del^{\mu} (g^{mn}) \Big) \nn\ .
\eea
Combining all the above, we finally obtain the following $d$-dimensional action
\beq
{\cal S} = \int \d^d x \sqrt{|g_{d}|} \left( \frac{M_p^2}{2} \R_d  - \frac{M_p^2}{2} \left( (d-2) \tau^{-2} (\del \tau)^2
+ \frac{D-d}{4} \rho^{-2} (\del \rho)^2 - \frac{1}{4}\, \del_{\mu} (g_{mn}) \del^{\mu} (g^{mn}) \right) - V \right) \ ,
\eeq
from which one can read the canonical fields. Before giving them, let us pursue with $g_{mn}$.

For our purposes, we will only consider here a single extra field $\sigma$. Its kinetic term was computed in \cite[App. B]{Andriot:2019wrs} for $D=10$ and $d=4$. For more $\sigma_I$, see \cite[App. D]{Andriot:2020wpp} and \cite{Andriot:2022yyj}. We recall the definition \cite{Danielsson:2012et}
\beq
g_{mn} \d y^m \d y^n  =  \sigma^A \delta_{ab} e^a e^b + \sigma^B \delta_{cd} e^c e^d \ , \label{sigmadef}
\eeq
where $a,b$ are parallel to an $O_p/D_p$ set, and $c,d$ transverse to it. We also have $e^a= e^a{}_m (y) \d y^m$. The exponents $A,B$ are chosen in such way that the determinant $g_{D-d}$ is independent of $\sigma$, and thus of coordinates $x$, as assumed above. This gives here
\beq
A (p+1-d) + B (D-p-1) = 0\ \rightarrow\ A=p+1-D , \ B =p+1-d \ .
\eeq
We then compute
\beq
- \frac{1}{4}\, \del_{\mu} (g_{mn}) \del^{\mu} (g^{mn}) = \frac{1}{4} AB(A-B) \sigma^{-2} (\del \sigma)^2 = \frac{1}{4} (D-p-1)(p+1-d)(D-d) \sigma^{-2} (\del \sigma)^2 \ .\nn
\eeq
We conclude by introducing the canonical fields
\begin{empheq}[box=\boxed]{align}
\label{eq:canonnormfields}
& \hat{\tau}= \sqrt{d-2}\, M_p\, \ln \tau \, ,\nn\\
& \hat{\rho}= \sqrt{\frac{D-d}{4}}\, M_p\, \ln \rho \, ,\\
& \hat{\sigma}= \sqrt{\frac{-AB(B-A)}{4}}\, M_p\, \ln \sigma = \sqrt{\frac{(D-p-1)(p+1-d)(D-d)}{4}}\, M_p\, \ln \sigma \, .\nn
\end{empheq}

\subsubsection{Scalar potential for $(\rho, \tau, \sigma)$}

We now consider $D=10$ and take as a starting point the 10d type II supergravities action (conventions of \cite[App. A]{Andriot:2016xvq}). The derivation of the scalar potential for $\rho,\tau$ was first done in \cite{Hertzberg:2007wc, Silverstein:2007ac} and matches the procedure to the above potential \eqref{pot0}. Let us extend this derivation here to arbitrary $d$, considering each term of the 10d action. We have first seen above that $\R_{10}$ would lead to the potential term $-\tau^{-2} \rho^{-1} \R_{10-d}$, where the $-\tau^{-2}$ comes from the prefactors of the action. Similarly, from 10d action terms, one gets the following potential terms in $d$ dimensions
\bea
& -\frac{1}{2} |H|^2 \ \rightarrow\ \frac{1}{2} \tau^{-2} \rho^{-3} |H|^2 \ ,\nn\\
& e^{\phi} \frac{T_{10}^{(p)}}{p+1} \ \rightarrow\ -\tau^{-\frac{d+2}{2}} \rho^{\frac{2p-8-d}{4}} g_s \frac{T_{10}^{(p)}}{p+1} \ ,\label{potterms}\\
& - e^{2\phi} \frac{1}{2} |F_q|^2 \ \rightarrow\ \frac{1}{2} \tau^{-d} \rho^{\frac{10-d-2q}{2}} g_s^2 |F_q|^2 \ ,\nn
\eea
reproducing the results of \cite{VanRiet:2011yc}. We used the fact that for sources $O_p/D_p$ of each dimensionality $p$, $T_{10}^{(p)}$ is inversely proportional to the transverse volume. We also consider that $p+1 \geq d$, as it should to preserve the $d$-dimensional maximal symmetry. For the source terms as well as for the fluxes, we extracted $\rho$ from the squares or the transverse volume, since it should be considered as a fluctuation around a vacuum valued field (here $g_{mn}$). We also considered, in agreement with our notations, the fluxes that are entirely along the $10-d$ compact dimensions (requiring $q \leq 10-d$): if it is not the case, the flux would vanish due to maximal symmetry of the $d$-dimensional spacetime, as long as $d>q$. A last contribution can come from spacetime-filling fluxes, possible when $d \leq q$. This contribution is rarely derived in detail, and we now turn to it.

Spacetime-filling fluxes in $d \geq 3$ come from $F^{10}_{3,4,5}$ and $H^{10}$: they were introduced (on-shell) in \eqref{Fqd} and \eqref{Fqdsquare} through the Hodge dual of the internal forms $F_{5,6,7}$ and $H_7$. On-shell actions with spacetime filling fluxes, as considered here for the scalar potential with such background flux contributions, require an extra $d$-dimensional boundary term. It allows to avoid an ill-defined boundary condition on a gauge potential (e.g.~$C_3|_{\infty}=0$), which would be gauge dependent, and trades it for a gauge independent condition ($\delta *_4 F_4 |_{\infty}=0$): we refer to \cite[App. A]{Andriot:2020lea} where this was detailed for $d=4$. The net result of this extra boundary term is that replaces a contribution $\varphi |F_4|^2$ to the potential by $-\varphi^{-1} |F_4|^2$. Starting as above from the 10d action terms with spacetime-filling fluxes (see \eqref{Fqd}, \eqref{Fqdsquare}), we end-up with the following potential terms
\bea
& \frac{1}{2} |H^d|^2 \ \rightarrow\ -\frac{1}{2} \tau^{-2} \tau^{2d} \rho^{d-3} |H_7|^2 \ \rightarrow\ \frac{1}{2} \tau^{2(1-d)} \rho^{3-d} |H_7|^2 \ ,\label{pottermsfilling}\\
& e^{2\phi} \frac{1}{2} |F^d_q|^2 \ \rightarrow\ -\frac{1}{2} \tau^{-2} \tau^{-(d-2)} \rho^{\frac{10-d}{2}} \tau^{2d} \rho^{d-q} g_s^2 |F_{10-q}|^2 \ \rightarrow\ \frac{1}{2} \tau^{-d} \rho^{\frac{10-d-2(10-q)}{2}} g_s^2 |F_{10-q}|^2 \ .\nn
\eea
In the second step, we separate each scalar contribution for clarity. We also note that $F_q^{10}$ is only a flux and does not contain any metric, despite the notations of \eqref{Fqd}, so the only $\rho,\tau$ contributions come from prefactors, $\phi$, and metric in the square. Finally, we use that on-shell, one has $|*_{10-d} F_{10-q}|^2 = |F_{10-q}|^2$. The last step implements the effect of the $d$-dimensional boundary term as previously described.

This allows us to give the precise scalar potential in $d$ dimensions, completing \eqref{pot0}
\begin{equation}
\label{pot}
\boxed{
  \begin{aligned}
V(\rho,\tau)= \frac{1}{2\kappa_{10}^2} \int \d^{10-d} y \sqrt{|g_{10-d}|}\ g_s^{-2}\ \Bigg( & \tau^{-2} \left(-\rho^{-1} \R_{10-d} + \frac{1}{2} \rho^{-3}\, |H|^2 \right) + \frac{1}{2} \tau^{2(1-d)} \rho^{3-d} |H_7|^2 \\
 & \hspace{-1.1in} -\, \tau^{-\frac{d+2}{2}}\, \sum_p \rho^{\frac{2p-8-d}{4}}\, g_s \frac{T_{10}^{(p)}}{p+1} + \frac{1}{2} \tau^{-d}\, g_s^2 \sum_{q=0}^{10-d} \rho^{\frac{10-d-2q}{2}}\, |F_q|^2  \Bigg) \ ,
 \end{aligned}
}
\end{equation}
where we recall that $H_7 = 0$ for $d\geq 4$.\\

We now focus on the dependency on $\sigma$, introduced around  \eqref{sigmadef}. The general potential for a single field $\sigma$ was derived in \cite{Andriot:2018ept}, together with \cite[App. A]{Andriot:2020lea} for the spacetime-filling fluxes, and concisely summarized in \cite[Sec. 2.2.1]{Andriot:2020lea}. We restrict ourselves here to a single $\sigma$ corresponding to a single source set of $O_p/D_p$ with contribution $T_{10}$; for multiple sets and multiple dimensionalities $p$, we refer to \cite{Andriot:2019wrs,Andriot:2022yyj}. We extend here the derivation of the potential for $\sigma$ to arbitrary $d$. It requires the notation $F_q^{(n)}$, referring to the part of the flux form with $n$ legs along the source set of $O_p/D_p$. For arbitrary $d$, one then gets the following fluctuations with respect to $\sigma$
\bea
& |H^{(n)}|^2, |F_q^{(n)}|^2 \ \rightarrow \ \sigma^{-An-B(q-n)} \times |H^{(n)}|^2, |F_q^{(n)}|^2 \ ,\nn\\
& T_{10} \ \rightarrow \ \sigma^{-\frac{1}{2}B(9-p)} \, T_{10} \ ,\\
& - |(*_{10-d} H_7)^{(n)}|^2, |(*_{10-d} F_{10-q})^{(n)}|^2 \ \rightarrow \ -\sigma^{-An-B(q-d-n)} \times |(*_{10-d} H_7)^{(n)}|^2, |(*_{10-d} F_{10-q})^{(n)}|^2 \nn\\
& \phantom{- |(*_{10-d} H_7)^{(n)}|^2, |(*_{10-d} F_{10-q})^{(n)}|^2 \ }\quad \quad \rightarrow \ \sigma^{An+B(q-d-n)} \times |(*_{10-d} H_7)^{(n)}|^2, |(*_{10-d} F_{10-q})^{(n)}|^2 \ ,\nn
\eea
with $A=p-9, B =p+1-d$. As already noticed in $d=4$ in \cite[App. A]{Andriot:2020lea} for the spacetime-filling fluxes, further convenient rewritings can be made. Since there are $p+1-d =B$ internal directions wrapped by a set of $O_p/D_p$, one verifies that
\bea
& *_{10-d} F_{10-q}^{(n)} = (*_{10-d} F_{10-q})^{(n')} \ {\rm with}\ n'=p+1-d-n \\
\Rightarrow\quad  & \ An' + B(q-d-n') = - An - B(10-q-n) \ ,\nn
\eea
allowing a rewriting of spacetime-filling flux potential terms. This eventually leads to the following potential
\begin{equation}
 \label{pot2}
 \boxed{
  \begin{aligned}
V(\rho,\tau,\sigma)=\ \frac{1}{2\kappa_{10}^2} \int & \d^{10-d} y \sqrt{|g_{10-d}|}\ g_s^{-2}\ \Bigg( \tau^{-2} \left(-\rho^{-1} \R_{10-d}(\sigma) + \frac{1}{2} \rho^{-3}\sum_n \sigma^{-An-B(3-n)} |H^{(n)}|^2 \right) \\&
+ \frac{1}{2} \tau^{2-2d} \rho^{3-d} \sum_n \sigma^{-An-B(7-n)} |H_7^{(n)}|^2 - \tau^{-\frac{d+2}{2}}\, \rho^{\frac{2p-8-d}{4}} \, \sigma^{\frac{1}{2}B(p-9)}\, g_s \frac{T_{10}}{p+1} \\&
+ \tau^{-d}\, \frac{1}{2} g_s^2 \sum_{q=0}^{10-d} \rho^{\frac{10-d-2q}{2}} \sum_n \sigma^{-An-B(q-n)} |F_q^{(n)}|^2 \Bigg)\ .
 \end{aligned}
}
\end{equation}
Note that the $\rho$ and $\sigma$ dependency of the $T_{10}$ term can be written as $\rho^{\frac{A+B}{4}} \, \sigma^{\frac{AB}{2}}$. The dependency $\R_{10-d}(\sigma)$ is more involved. While the general result was provided in \cite{Andriot:2018ept}, we can restrict ourselves here to the case of compact group manifolds. Then, the expression of the Ricci scalar in terms of structure constants $f^a{}_{bc}$ is formally unchanged (only the dimension is), and can be expressed as
\beq
{\cal R}_{10-d} =  {\cal R}_{||} +  {\cal R}_{||}^{\bot}  - \frac{1}{2} |f^{{}_{||}}{}_{{}_{\bot} {}_{\bot}}|^2  - \delta^{cd}   f^{b_{\bot}}{}_{a_{||}  c_{\bot}} f^{a_{||}}{}_{ b_{\bot} d_{\bot}}  \ , \label{R6final2}
\eeq
where we refer to \cite{Andriot:2018ept, Andriot:2020lea} for the detailed expression of each term. The fluctuation with respect to $\sigma$ is also formally the same, simply with the new values of $A,B$, and can be phrased as
\beq
{\cal R}_{10-d} (\sigma) = -\sigma^{-B}\ (\delta^{cd}   f^{b_{\bot}}{}_{a_{||}  c_{\bot}} f^{a_{||}}{}_{ b_{\bot} d_{\bot}})^0 + \sigma^{-A} \left( {\cal R}_{||} +  {\cal R}_{||}^{\bot} \right)^0  - \frac{1}{2} \sigma^{-2B+A} |f^{{}_{||} 0}{}_{{}_{\bot} {}_{\bot}}|^2  \ ,\label{R6sigmaN3}
\eeq
and we will drop the background labels ${}^0$. This ends the derivation of the scalar potential.\\

In the following, we will use the simplified notation
\beq
\frac{\int \d^{10-d} y \sqrt{|g_{10-d}|}\ T_{10} }{\int \d^{10-d} y \sqrt{|g_{10-d}|}}\ \rightarrow\ T_{10}\ , \label{simpnot}
\eeq
for each internal quantity (such as here $T_{10}$) entering the potential. This notation is an equality in the case where these quantities are constant. This notation allows to replace the prefactor in $V$ by $\frac{M_p^2}{2}$.

\subsubsection{Kinetic terms and scalar potential for $(\tau, r)$}

We now turn to a different set of fields: the 4d dilaton $\tau$ and the radion $r$. Those were introduced in \cite[Sec. 2.2.2]{Andriot:2020lea} and we generalize here the derivation to arbitrary dimensions. These fields enter the $D$-dimensional metric as follows
\beq
\d s_D^2=\tau^{-2}(x) g_{\mu \nu}(x) \d x^\mu \d x^\nu + r^2(x) (e^1)^2 + \sum_{a=2}^{D-d} \delta_{ab} e^a e^b \ ,
\eeq
where $e^a = e^a{}_m (y) \d y^m$. The $D$-dimensional Ricci scalar will then again lead to $\tau^{2} {\cal R}_d + ...$. To reach the $d$-dimensional Einstein frame, we then define the dilaton fluctuation as
\beq
e^{\delta \phi} = \tau^{\frac{2-d}{2}}\, r^{\frac{1}{2}} \ ,
\eeq
and one gets the same $-\tau^{-2}$ prefactor entering the potential as in \eqref{pot0}.

To compute the kinetic terms, one considers again the combination $ \tau^{-2} (\R_{D} + 4 (\del \phi)_{D}^2)$. Without loss of generality, we can take for this computation the $D$-dimensional metric to be given by $g_{\mu \nu}=\eta_{\mu \nu}$ and $e^a{}_m = \delta^a_m$. In addition we introduce the indices $i,j = 2, ..., D-d$, along which there is no scalar field. One then derives the following expression for the Ricci scalar
\bea
\R_D =\ &2 \tau^2 \del_\mu \Big((d-1) \del^\mu \ln \tau - \del^\mu \ln r \Big) - 2 \tau^2 \left(\del \ln r \right)^2 \nn \\&
- (d-2)(d-1) \left(\del \tau \right)^2 + 2(d-2)\tau^2 \del_\mu \ln \tau \, \del^\mu \ln r \ ,
\eea
and the dilaton contribution
\bea
4(\del \phi_D)^2 =\ (d-2)^2 \left(\del \tau \right)^2 + 2(2-d) \tau^2 \del_\mu \ln \tau \, \del^\mu \ln r + \tau^2 \left(\del \ln r \right)^2 \ .
\eea
Combining these contributions as in \eqref{Sdgen}, leaving aside the total derivative, we then get the following $d$-dimensional action
\beq
{\cal S} = \int \d^d x \sqrt{|g_d|} \left(\frac{M_p^2}{2} \R_d - \frac{M_p^2}{2} \left( (d-2) \left(\del \ln \tau \right)^2 + \left(\del \ln r \right)^2 \right) - V \right)
\eeq
from which we read the kinetic terms and the canonical fields $\hat{\tau}, \hat{r}$
\beq
\label{eq:NewPotCanFields}
\boxed{\hat{\tau} = \sqrt{d-2}\, M_p \ln \tau \ , \quad \hat{r}=M_p \ln r \ .}
\eeq

We now focus on $D=10$ and derive the scalar potential. We introduce a new notation for the fluxes: $F_q^{(n)}$ includes the components with $n$ legs along direction $1$ (so $n=0,1$). We then get the following potential terms starting with the 10d action ones
\bea
& -\frac{1}{2} |H|^2 \ \rightarrow\ \frac{1}{2} \tau^{-2} \left( |H^{(0)}|^2 + r^{-2} |H^{(1)}|^2 \right) \ ,\nn\\
& e^{\phi} \frac{T_{10}^{(p)}}{p+1} \ \rightarrow\ -\tau^{-\frac{d+2}{2}}  \left(\delta_1^\parallel r^{\frac{1}{2}} + \delta_1^\perp r^{-\frac{1}{2}} \right) g_s \frac{T_{10}^{(p)}}{p+1} \ ,\label{pottermsr}\\
& - e^{2\phi} \frac{1}{2} |F_q|^2 \ \rightarrow\ \frac{1}{2} \tau^{-d}  g_s^2  \left(r |F_q^{(0)}|^2 + r^{-1} |F_q^{(1)}|^2 \right) \ ,\nn
\eea
where the source term depends on whether $1$ is parallel or transverse to the source. The spacetime filling fluxes go under the same steps as described previously for $(\rho,\tau,\sigma)$, namely
\bea
& \frac{1}{2} |H^d|^2 \ \rightarrow\ -\frac{1}{2} \tau^{2d-2} \left( |(*_{10-d}H_7)^{(0)}|^2 + r^{-2}  |(*_{10-d}H_7)^{(1)}|^2 \right) \label{pottermsfillingr}\\
& \phantom{\frac{1}{2} |H^d|^2 \ } \quad \quad \rightarrow\ \frac{1}{2} \tau^{2(1-d)} \left( |(*_{10-d}H_7)^{(0)}|^2 + r^2 |(*_{10-d}H_7)^{(1)}|^2 \right) \ ,\nn\\
& e^{2\phi} \frac{1}{2} |F^d_q|^2 \ \rightarrow\ -\frac{1}{2} \tau^{d} g_s^2  \left( r |(*_{10-d}F_{10-q})^{(0)}|^2 +  r^{-1} |(*_{10-d}F_{10-q})^{(1)}|^2 \right) \nn\\
& \phantom{e^{2\phi} \frac{1}{2} |F^d_q|^2 \ } \quad \quad \rightarrow\ \frac{1}{2} \tau^{-d} g_s^2 \left( r^{-1} |(*_{10-d}F_{10-q})^{(0)}|^2 +  r |(*_{10-d}F_{10-q})^{(1)}|^2 \right) \ .\nn
\eea
The latter terms can be further simplified by noticing that on-shell, by definition of the Hodge star,
\beq
|(*_{10-d}H_7)^{(n)}|^2 = |H_7^{(|1-n|)}|^2 \ ,\ |(*_{10-d}F_{10-q})^{(n)}|^2 = |F_{10-q}^{(|1-n|)}|^2 \ .
\eeq
We eventually obtain the following potential
\begin{equation}
 \label{rpot2}
 \boxed{
  \begin{aligned}
V(\tau, r)=\ \frac{1}{2\kappa_{10}^2} \int & \d^{10-d} y \sqrt{|g_{10-d}|}\ g_s^{-2}\ \Bigg( \tau^{-2} \left(- \R_{10-d}(r) + \frac{1}{2} \left( |H^{(0)}|^2 + r^{-2} |H^{(1)}|^2 \right) \right) \\&
+ \frac{1}{2} \tau^{2-2d} \left( |H_7^{(1)}|^2 + r^2  |H_7^{(0)}|^2 \right)- g_s \tau^{-\frac{d+2}{2}} \sum_{p,I} \left(\delta_1^{\parallel_I} r^{\frac{1}{2}} + \delta_1^{\perp_I} r^{-\frac{1}{2}} \right) \frac{T_{10}^{(p)_I}}{p+1} \\&
+ \tau^{-d}\, \frac{1}{2} g_s^2\sum_{q=0}^{10-d} \left(r |F_q^{(0)}|^2 + r^{-1} |F_q^{(1)}|^2 \right) \Bigg)
 \end{aligned}
}
\end{equation}
In the following, we will make use of the above in the case of a group manifold, for which the Ricci scalar dependency on $r$ can be written as follows
\beq
\R_{10-d}(r)=\R_{10-d}^0 + (r^{-2}-1)\R_{11}^0 + \frac{1}{4} (2-r^2-r^{-2}) \delta^{ik} \delta^{jl} {{f^1}_{ij}}^0 {{f^1}_{kl}}^0 \ .
\eeq
It is indeed independent of the dimension $d$, so we can use this formula obtained in \cite[Sec. 2.2.2]{Andriot:2020lea} for $d=4$.

\subsection{No-go theorems in arbitrary dimension $d$, and $c$ values}\label{sec:dnogo}

Using the $d$-dimensional effective theories worked-out in Section \ref{sec:dimred}, we now reproduce the no-go theorems of Sections \ref{sec:MN}-\ref{sec:Het} obtained using 10d equations: the corresponding no-go theorems in $d$ dimensions ($10 \geq d \geq 3$) are given here in Sections \ref{sec:MNd}-\ref{sec:Hetd}. For each no-go theorem, we refer to Sections \ref{sec:MN}-\ref{sec:Het} for literature references. In addition to those, we give two further no-go theorems in Section \ref{sec:lambda} and \ref{sec:InternEinstein}: obtaining those in 10d would require to use the internal Einstein equation, or partial traces thereof, which we have not introduced in this paper.

The $d$-dimensional formulation of the no-go theorems takes the form of an inequality
\beq
\label{eq:nogolincomb}
a\, V + \sum_{i} b_i\, \varphi^i \del_{\varphi^i} V  \leq 0
\eeq
with $a>0, \exists\, b_i \neq 0$, valid upon some assumption. By definition, such an inequality indeed forbids de Sitter solutions, which correspond to critical points of the potential. Thanks to the relation \eqref{RdV} between $V$ and ${\cal R}_d$, one can verify that, at a critical point where fields $\varphi^i=1$, the left-hand side of the inequality matches the 10d equations providing the (same) no-go theorem. A related discussion can be found in Section \ref{sec:quasidS}.

We then rewrite the inequality in terms of canonical fields, using \eqref{eq:canonnormfields} for $(\rho,\tau,\sigma)$ or \eqref{eq:NewPotCanFields} for $(\tau, r)$, in order to deduce the $c$ value as shown in \cite{Andriot:2019wrs,Andriot:2021rdy}
\beq
\label{eq:NoGoEq}
a\, V + \sum_i M_p\, \hat{b}_{i}\, \del_{\hat{\varphi}^i}V \le 0 \ \ \Rightarrow  \ \ |\nabla V| \geq \frac{c}{M_p}\, V\ , \quad  c^2= \frac{a^2}{\sum_i \hat{b}_{i}^2} \ .
\eeq
Note that the $M_p$ appearing above will be dropped in the following to alleviate the equations. The constant $c$ corresponds to the one of the swampland de Sitter conjecture \cite{Obied:2018sgi}. Of particular interest here is its comparison, across dimensions, to the TCC bound \cite{Bedroya:2019snp}
\beq
c_0 = \frac{2}{\sqrt{(d-1)(d-2)}} \ . \label{TCC}
\eeq
This will be analysed in Section \ref{sec:discussionC}. In $d=4$, the $c$ values for these no-go theorems and more were obtained in \cite{Andriot:2020lea}; here, we only consider those with field independent assumptions.

\subsubsection{Extension of Maldacena-Nuñez}\label{sec:MNd}

The no-go theorem is obtained as follows, matching \eqref{eq:dRicciScalar2}
\bea
& \frac{2}{M_p^2} \left( 2 V + \tau \del_\tau V \right) \\
& = \frac{d-2}{2} \left(\tau^{-\frac{d+2}{2}} \sum_p\, \rho^{\frac{2p-8-d}{4}} g_s \frac{T_{10}^{(p)}}{p+1} - \tau^{-d} g_s^2 \sum_{q=0}^{10-d} \rho^{\frac{10-d-2q}{2}}\, |F_q|^2 - 2 \tau^{2-2d} \rho^{3-d} |H_7|^2  \right) \nn \\
& \Rightarrow \quad  2V + \sqrt{d-2}\, \del_{\hat{\tau}} V \le 0 \quad \text{if all $T_{10}^{(p)} \le 0$.} \nn\\
&  \Rightarrow \quad  c^2 = \frac{4}{(d-2)} \nn
\eea

\subsubsection{No-go for $p=7,8,9$, or $p=4,5,6$ with $F_{6-p}=0$, or $p=2$ with $H=0$}\label{sec:NonVanFluxd}

\begin{figure}[H]
\centering
\includegraphics[width=0.7\textwidth]{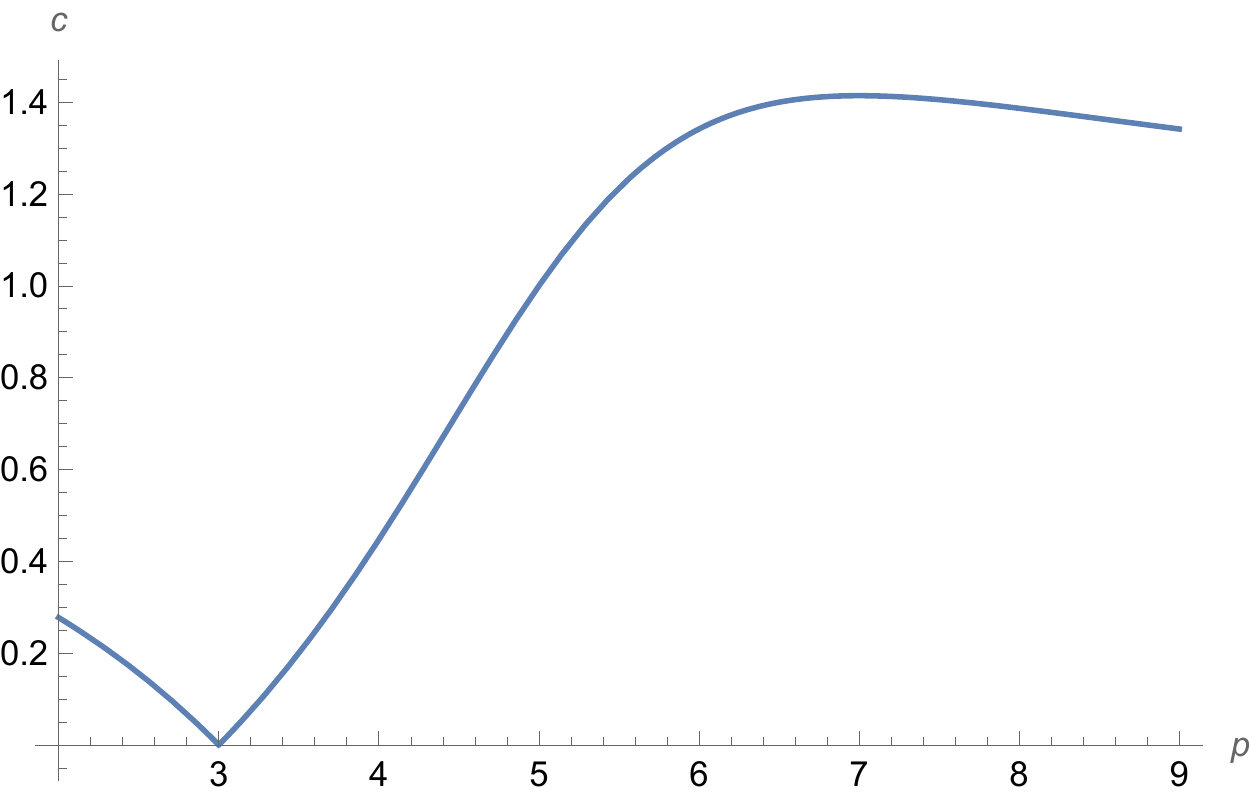}
\caption{Graph of $c(p,d=6)$ as obtained in \eqref{nogo789d}. The behaviour of this function is qualitatively the same for any dimension $2<d<10$.}\label{fig:cp}
\end{figure}

A first part of the no-go theorem is obtained as follows, matching \eqref{eq:singlesizecomb}
\bea
& \frac{2}{M_p^2} \left( \frac{2(p-3)}{d-2} V - \frac{d+4-2p}{2(d-2)} \tau \del_\tau V + \rho \del_\rho V \right) \label{nogo789d}\\
& = - \tau^{-2} \rho^{-3} |H|^2 + (4-p) \tau^{2-2d} \rho^{3-d} |H_7|^2 + \tau^{-d}\frac{1}{2} g_s^2 \sum_{q=0}^{10-d} \rho^{\frac{10-d-2q}{2}} \left( 8-p-q \right) |F_q|^2 \nn \\
& \Rightarrow \quad \frac{2(p-3)}{d-2} V - \frac{d+4-2p}{2\sqrt{d-2}} \del_{\hat{\tau}} V + \sqrt{\frac{10-d}{4}} \del_{\hat{\rho}} V \le 0 \quad \text{if $p=7,8,9$, or $p=4,5,6$ \& $F_{6-p} = 0$.} \nn\\
& \Rightarrow \quad  c^2 = \frac{4 (p-3)^2}{(d-2)(5d-1-4p-pd+p^2)} = \frac{4 (p-3)^2}{(d-2)((p-3)^2+(p-5)(2-d))} \nn\\
& \phantom{\Rightarrow \quad  c^2 = \frac{4 (p-3)^2}{(d-2)(5d-1-4p-pd+p^2)}}\ = \frac{4}{d-2} + \frac{4 (p-5)}{(p-3)(p-1-d)+2 (d-2)} \nn
\eea
The denominators appearing in the formulas for $c^2$ do not vanish for any $p$ with $2<d<10$. However, they do admit a (single) root at $d=10$, namely $p=7$: this can also be seen from the coefficients of the linear combination. Nevertheless, $p=7$ is not allowed for $d=10$ since the minimal $p$ allowed by maximal symmetry is $p=d-1$. Keeping the latter in mind, the formula for $c$ is then always valid for our purposes. It is then straightforward to analyse the dependency of $c$ on $p$, especially using the last formula for $c^2$. Interestingly, it gets minimized at a value of $p$ independent of $d$, namely $p=3$ (and maximized at $p=7$): see Figure \ref{fig:cp}. The first part of the no-go however requires as an assumption that $4 \leq p \leq 9$. In that range, the minimal value of $c$ is obtained at $p=4$ (as in $d=4$). We conclude
\beq
c^2 = \frac{4 (p-3)^2}{(d-2)(5d-1-4p-pd+p^2)} \geq \frac{4}{(d-2)(d-1)} \ ,\quad (9 \geq p\geq 4)
\eeq
with saturation at $p=4$.

If we now divide \eqref{nogo789d} by $p-3$ and focus on $p=2$ (implicitly restricting us to $d=3$), we get the second part of the no-go theorem assuming $H=0$, as noticed in 10d in \eqref{eq:p2nogo}. The value for $c$ is then the same as above and becomes for $p=2$
\beq
c^2 =  \frac{4}{(d-2)(3d-5)} < \frac{4}{(d-2)(d-1)}  \ ,\quad (p=2)  \label{cd=3p=2}
\eeq
which strictly speaking should be evaluated at $d=3$. For the first time, we get a value for $c$ which is lower than the TCC bound. This can be seen on Figure \ref{fig:cp}, comparing to $p=4$. This highlights the peculiarity of $d=3$; we will come back to this point in Section \ref{sec:discussionC}.

\subsubsection{Positive or vanishing internal curvature $\R_{10-d}$}\label{sec:PosCurvd}

The no-go theorem is obtained as follows, matching \eqref{eq:singlesizecomb2}
\bea
& \frac{2}{M_p^2} \left( \frac{2(d+p-1)}{d-2} V - \frac{d-4-2p}{2(d-2)} \tau \del_\tau V + \rho \del_\rho V \right) \\
& = - 2 \tau^{-2} \rho^{-1} \R_{10-d} + (1-p)  \tau^{2-2d} \rho^{3-d} |H_7|^2 + \frac{1}{2} \tau^{-d} g_s^2 \sum_{q=0}^{10-d} \rho^{\frac{10-d-2q}{2}}\, \left( 6-p-q \right) |F_q|^2 \nn \\
& \Rightarrow \quad  \frac{2(d+p-1)}{d-2} V - \frac{d-4-2p}{2\sqrt{d-2}} \del_{\hat{\tau}} V + \sqrt{\frac{10-d}{4}} \del_{\hat{\rho}} V \le 0 \quad \text{if $\R_{10-d} \ge 0$ and $p \ge 4$.} \nn\\
& \Rightarrow \quad  c^2 = \frac{4(d+p-1)^2}{(d-2)(-1+d-dp+4p+p^2)} \nn
\eea
As above, the denominator of $c^2$ does not vanish for $2<d <10$, but it does at $d=10$ for $p=3$. However, the latter $p$ value is not allowed in $d=10$, so the formula for $c$ is again always valid for our purposes. We can then analyse the dependency of $c$ in $p$. One can prove that $c^2 > 1$ within the admissible range of our parameters, as in $d=4$. The proof is not especially enlightening and we refrain from detailing it here.

\subsubsection{No-go for $p=d-1$}\label{sec:d-1d}

The no-go theorem is obtained as follows, matching \eqref{eq:riccinogo2}
\bea
& \frac{2}{M_p^2} \Bigg( \frac{2(p+1)}{d-2} V - \frac{d-4-2p}{2(d-2)} \tau \del_\tau V + \rho \del_\rho V \Bigg) \\
& = - \left|\tau^{-1} \rho^{-\frac{3}{2}} *_{10-d} H + \varepsilon_p g_s \tau^{-\frac{d}{2}} \rho^{\frac{2p-2-d}{4}} F_{6-p} \right|^2 + 2  \tau^{-\frac{d+2}{2}} \rho^{\frac{2p-8-d}{4}} g_s \varepsilon_p (\d F_{8-p}) \nn \\
& \ +\frac{1}{2} \tau^{-d} g_s^2 \displaystyle \sum_{\substack{q=0 \\ q\neq 6-p}}^{10-d} \rho^{\frac{10-d-2q}{2}}\, |F_q|^2 \left(4-p-q\right) - p \tau^{2-2d} \rho^{3-d} |H_7|^2 \nn
\eea
where we have used the sourced Bianchi identity \eqref{BIscalar} multiplied by scalar fields and rearranged them conveniently for the off-shell formulation of the no-go theorem. All terms on the right-hand side are negative for $d\geq 4$, or for $d=3$ with $F_0=0$, except the one in $(\d F_{8-p})$ that has a priori no definite sign. However, with the same argumentation as in Section \ref{sec:d-1}, it follows from equation \eqref{eq:MaxFluxCond} that the integral of this term vanishes for $p=d-1$, since it is a total derivative over the compact space. Here, the integral is actually implicit thanks to our simplifying notations \eqref{simpnot}, so this term on the right-hand side simply vanishes. We obtain
\bea
&\frac{2d}{d-2} V + \frac{d+2}{2\sqrt{d-2}} \del_{\hat{\tau}} V + \sqrt{\frac{10-d}{4}} \del_{\hat{\rho}} V \le 0 \quad \text{if $p = d-1$, and $d\geq 4$ or $d=3$ with $F_0=0$.} \nn\\
& \Rightarrow \quad c^2 = \frac{d^2}{(d-2)(d-1)}
\eea

\subsubsection{Heterotic at order $(\alpha')^0$}\label{sec:Hetd}

The (bosonic) action for heterotic strings at order $(\alpha')^0$ reduces to the NSNS part, so the potential \eqref{pot} boils down to the $\R_{10-d}$ and $H$-flux terms. The no-go theorem is then obtained as follows, matching \eqref{het10d}
\bea
& \frac{2}{M_p^2} \left( V + \frac{4-d}{4} \tau \del_\tau V + \frac{d-2}{2} \rho \del_\rho V \right) = -\frac{d-2}{2} \tau^{-2} \rho^{-3} |H|^2 \\
& \Rightarrow \quad V + \frac{(4-d)\sqrt{d-2}}{4} \del_{\hat{\tau}} V + \frac{(d-2)\sqrt{10-d}}{4} \del_{\hat{\rho}} V \le 0 \nn\\
& \Rightarrow \quad  c^2 = \frac{4}{(d-2)(d-1)} \nn
\eea

\subsubsection{The requirement $\lambda > 0$}\label{sec:lambda}

This no-go theorem was first derived in 4d in \cite[(3.12)]{Andriot:2018ept}, and $c$ was computed in $d=4$ in \cite[Sec. 3.1]{Andriot:2019wrs} and \cite{Andriot:2020lea}. It assumes more restrictions than previously on the ansatz, namely that we work on a group manifold with constant flux components. The latter allows to restrict the $H^{(n)}$ components to $n=0,2$, because it must get a minus sign under any orientifold involution. In addition, being on a group manifold with constant flux components restricts the components entering the RR sourced Bianchi identity to $F_{8-p}^{(1)}$ and $F_{6-p}^{(0)}$. We then consider the following combination of the potential \eqref{pot2} (with curvature \eqref{R6sigmaN3}) and its derivatives
\bea
& \frac{2}{M_p^2}  \Bigg( \frac{4(B-A)}{(d-2)} V +\frac{(B-A)(d+2)}{2(d-2)} \tau \del_\tau V - (A+B) \rho \del_\rho V + 2 \sigma \del_\sigma V \Bigg) \\
& \qquad = (B-A) \Bigg( - 2 \tau^{-2} \rho^{-1} \left( \sigma^{-B} \delta^{cd} {f^{b_\perp}}_{a_\parallel c_\perp} {f^{a_\parallel}}_{b_\perp d_\perp} + \sigma^{A-2B} |f^{{}_{||}}{}_{{}_{\bot} {}_{\bot}}|^2 \right) \nn \\
& \qquad \phantom{=(A-B) \Bigg(} - 2 \tau^{-2} \rho^{-3} \sigma^{-3B} |H^{(0)}|^2  + (n-1-p)  \tau^{2-2d} \rho^{3-d} \sum_n \sigma^{-An-B(7-n)} |H_7^{(n)}|^2 \nn \\
& \qquad \phantom{=(A-B) \Bigg(} +4 \tau^{-\frac{d+2}{2}}\, \rho^{\frac{2p-8-d}{4}} \, \sigma^{\frac{1}{2}B(p-9)}\, g_s \frac{T_{10}}{p+1}   \nn \\
& \qquad \phantom{=(A-B) \Bigg(} + \frac{1}{2} \tau^{-d} g_s^2 \sum_{q=0}^{10-d} \rho^{\frac{10-d-2q}{2}} \sum_n \sigma^{-An-B(q-n)} ( 2+2n-p-q ) |F_q^{(n)}|^2 \Bigg) \nn
\eea
We now make use of the sourced Bianchi identity \eqref{BIscalar} (for arbitrary $p$) multiplied by scalar fields as in Section \ref{sec:d-1d}. We complete the reformulation of this Bianchi identity with the following rewriting of the $(\d F_{8-p})_\perp$ term (we recall that the only relevant component here is $F_{8-p}^{(1)}$)
\bea
2 \tau^{-\frac{d+2}{2}} \rho^{\frac{2p-8-d}{4}} & \sigma^{B\frac{p-9}{2}} \varepsilon_p g_s (\d F_{8-p})_\perp \nn \\
=\ & -\sum_{a_\parallel} \left|\tau^{-1} \rho^{-\frac{1}{2}} \sigma^{\frac{A-2B}{2}} *_\perp \left(\d e^{a_\parallel}\right)|_\perp - \tau^{-\frac{d}{2}} \rho^{\frac{2p-6-d}{4}} \sigma^{\frac{-A-B(7-p)}{2}} \varepsilon_p g_s \left(\iota_{\del_{a_\parallel}} F_{8-p}^{(1)} \right)\right|^2 \nn \\
& +\tau^{-d} \rho^{p-3-\frac{d}{2}} \sigma^{-A-B(7-p)} g_s^2 |F_{8-p}^{(1)}|^2 + \tau^{-2} \rho^{-1} \sigma^{A-2B} |{f^{\parallel}}_{\perp \perp}|^2 \ .\label{BIscalar2}
\eea
We then obtain
\bea
& \frac{2}{M_p^2}  \Bigg( \frac{4(B-A)}{(d-2)} V +\frac{(B-A)(d+2)}{2(d-2)} \tau \del_\tau V - (A+B) \rho \del_\rho V + 2 \sigma \del_\sigma V \Bigg) \label{eq:PotCombLambda}\\
& = (B-A) \Bigg( - 2 \tau^{-2} \rho^{-1} \sigma^{-B} \delta^{cd} {f^{b_\perp}}_{a_\parallel c_\perp} {f^{a_\parallel}}_{b_\perp d_\perp}  + (n-1-p)  \tau^{2-2d} \rho^{3-d} \sum_n \sigma^{-An-B(7-n)} |H_7^{(n)}|^2  \nn \\
& \phantom{=(A-B) \Bigg(} - 2 \left|\tau^{-1} \rho^{-\frac{3}{2}} \sigma^{\frac{-3}{2}B} *_\perp H^{(0)} + \varepsilon_p g_s \tau^{-\frac{d}{2}} \rho^{\frac{2p-2-d}{4}} \sigma^{B \frac{p-6}{2}} F_{6-p}^{(0)}\right|^2  \nn \\
& \phantom{=(A-B) \Bigg(} -2 \sum_{a_\parallel} \left|\tau^{-1} \rho^{-\frac{1}{2}} \sigma^{\frac{A-2B}{2}} *_\perp \left(\d e^{a_\parallel}\right)|_\perp - \tau^{-\frac{d}{2}} \rho^{\frac{2p-6-d}{4}} \sigma^{\frac{-A-B(7-p)}{2}} \varepsilon_p g_s \left(\iota_{\del_{a_\parallel}} F_{8-p}^{(1)} \right)\right|^2  \nn \\
& \phantom{=(A-B) \Bigg(} + \frac{1}{2} \tau^{-d} g_s^2 \displaystyle \sum_{\substack{ q,n \\ q\neq 6-p, (n=0) \\ q\neq 8-p, (n=1)}} \rho^{\frac{10-d-2q}{2}}  \sigma^{-An-B(q-n)}\, ( 2+2n-p-q )\, |F_q^{(n)}|^2 \Bigg) \nn
\eea
Since $H_7$ is non-zero only in $d=3$, where it is proportional to the internal volume, it is then along all internal directions wrapped by the sources. Therefore, it must have a single value $n=p+1-d$. This implies that $n-1-p = - d < 0$. Similarly, the coefficient of RR fluxes is negative: indeed, on the one hand $n \leq q$ and on the other hand $n \leq p+1-d$, since $n$ is the number of (internal) legs along the sources. Together this leads to $2+2n -p-q \leq 3-d \leq 0$. We conclude that the parentheses is negative if
\beq
-\delta^{cd} {f^{b_\perp}}_{a_\parallel c_\perp} {f^{a_\parallel}}_{b_\perp d_\perp} \equiv \lambda \, |{f^\parallel}_{\perp \perp}|^2 \leq 0 \ .
\eeq
Given the requirement for de Sitter to have $|{f^\parallel}_{\perp \perp}|^2 \neq 0$ in $d=4$ \cite{Andriot:2016xvq}, we shall with some abuse refer to the previous condition as $\lambda \leq 0$. Since $B-A= 10-d$, we conclude for $3\leq d \leq 9$
\bea
& \hspace{-0.2in }\frac{4(B-A)}{(d-2)} V + \frac{(B-A)(d+2)}{2\sqrt{d-2}} \del_{\hat{\tau}} V - (A+B) \sqrt{\frac{10-d}{4}} \del_{\hat{\rho}} V + \sqrt{-AB(B-A)} \del_{\hat{\sigma}} V \le 0 \\
& \hspace{4.2in} \text{if $3\leq d \leq 9$ and $ \lambda \le 0$.} \nn\\
& \hspace{-0.2in } \Rightarrow \qquad c^2 = \frac{4}{(d-2)(d-1)}  \nn
\eea

\subsubsection{The internal parallel Einstein equation}\label{sec:InternEinstein}

We generalize here to arbitrary dimension $d$ a no-go theorem obtained using 10d equations in \cite[Sec. 3.2]{Andriot:2019wrs} and $d=4$ effective potential in \cite[Sec. 2.3]{Andriot:2020lea}. We consider the following combination of the potential $V(\tau,r)$ given in \eqref{rpot2} and its derivatives
\bea
& \frac{2}{M_p^2} \left( 2V + \tau \del_\tau V + (d-2) r \del_r V \right)\\
& = - (d-2) \left(\tau^{-2} r^{-2} |H^{(1)}|^2 + \tau^{2-2d} |H_7^{(1)}|^2 + \tau^{-d} r^{-1} g_s^2 \sum_q  |F_q^{(1)}|^2  \right) \nn \\
&\ \ - (d-2) \left(- g_s \tau^{-\frac{d+2}{2}} r^{-\frac{1}{2}} \sum_{p,I} \frac{T_{10}^{(p)_I}}{p+1} \delta_1^{\bot_I} + \tau^{-2} r^{-2} \left( \frac{1}{2} \delta^{ik} \delta^{jl} {{f^1}_{ij}} {{f^1}_{kl}}  - 2\R_{11} \right) \right) \nn\\
&\ \ - (d-2) \left( - \tau^{-2} r^2  \frac{1}{2} \delta^{ik} \delta^{jl} {{f^1}_{ij}} {{f^1}_{kl}}  \right) \nn
\eea
As shown in \cite[(2.66)]{Andriot:2020lea}, $ \frac{1}{2} \delta^{ik} \delta^{jl} {{f^1}_{ij}} {{f^1}_{kl}}  - 2\R_{11} \geq 0 $ on a group manifold. If we now restrict ourselves to the radion direction $1$ being parallel to all source sets $I$ with $O_p$ (i.e.~those are wrapped along $1$; this requires $p\geq d$), implying $\delta_1^{\perp_I} = 0$, then only the last term has the wrong sign. Note that sets with only $D_p$ do not matter since they obey $T_{10}^{(p)_I}<0$. We conclude
\bea
\label{eq:IntParEinComb}
& 2 V + \sqrt{d-2}\, \del_{\hat{\tau}} V + (d-2)\, \del_{\hat{r}} V \le 0 \quad \text{if ${f^1}_{ij} =0$ for direction $1$ being parallel and $\forall \ i,j$.} \nn \\
& \Rightarrow \quad c^2 = \frac{4}{(d-2)(d-1)}
\eea
where canonical fields are given in \eqref{eq:NewPotCanFields}.

\section{Discussion: swampland conjectures and accelerated expansion in $d$ dimensions}\label{sec:discussionC}

In Section \ref{sec:dnogo}, we have derived 7 no-go theorems against a (quasi-) de Sitter spacetime in $d$ dimensions, $3 \leq d \leq 10$, in a $d$-dimensional theory coming from a compactification of 10d type IIA/B supergravity compactifications. The 5 first no-go theorems were also derived in Section \ref{sec:10dnogo} using 10d equations. The no-go theorems in Section \ref{sec:dnogo} are valid upon some assumptions, which are always taken to be field independent: they only depend on the 10d background. The work of Section \ref{sec:dnogo} extends to arbitrary $d$ dimensions the study of \cite{Andriot:2020lea} done for $d=4$. For each of the no-go theorems, we have derived the ($d$-dependent) $c$ value, as defined in \eqref{eq:NoGoEq}. It is meant to correspond to the $c$ of the de Sitter swampland conjecture \cite{Obied:2018sgi}, and we are particularly interested in the comparison to the TCC bound \eqref{TCC}, a $c$ value obtained in \cite{Bedroya:2019snp} using the TCC, a refined version of the de Sitter conjecture.

We first report in Section \ref{sec:cvalues} on the $c$ values obtained and compare them to the TCC bound. The case of $d=3$ will be singled-out. We then compare in Section \ref{sec:swampconj} to other values proposed in the literature, and discuss further related conjectures. We finally say a word on a related bound for cosmic accelerated expansion in Section \ref{sec:accexp}.

\subsection{$c$ values obtained, and the $d=3$ case}\label{sec:cvalues}

\begin{figure}[h]
\centering
\includegraphics[width=\textwidth]{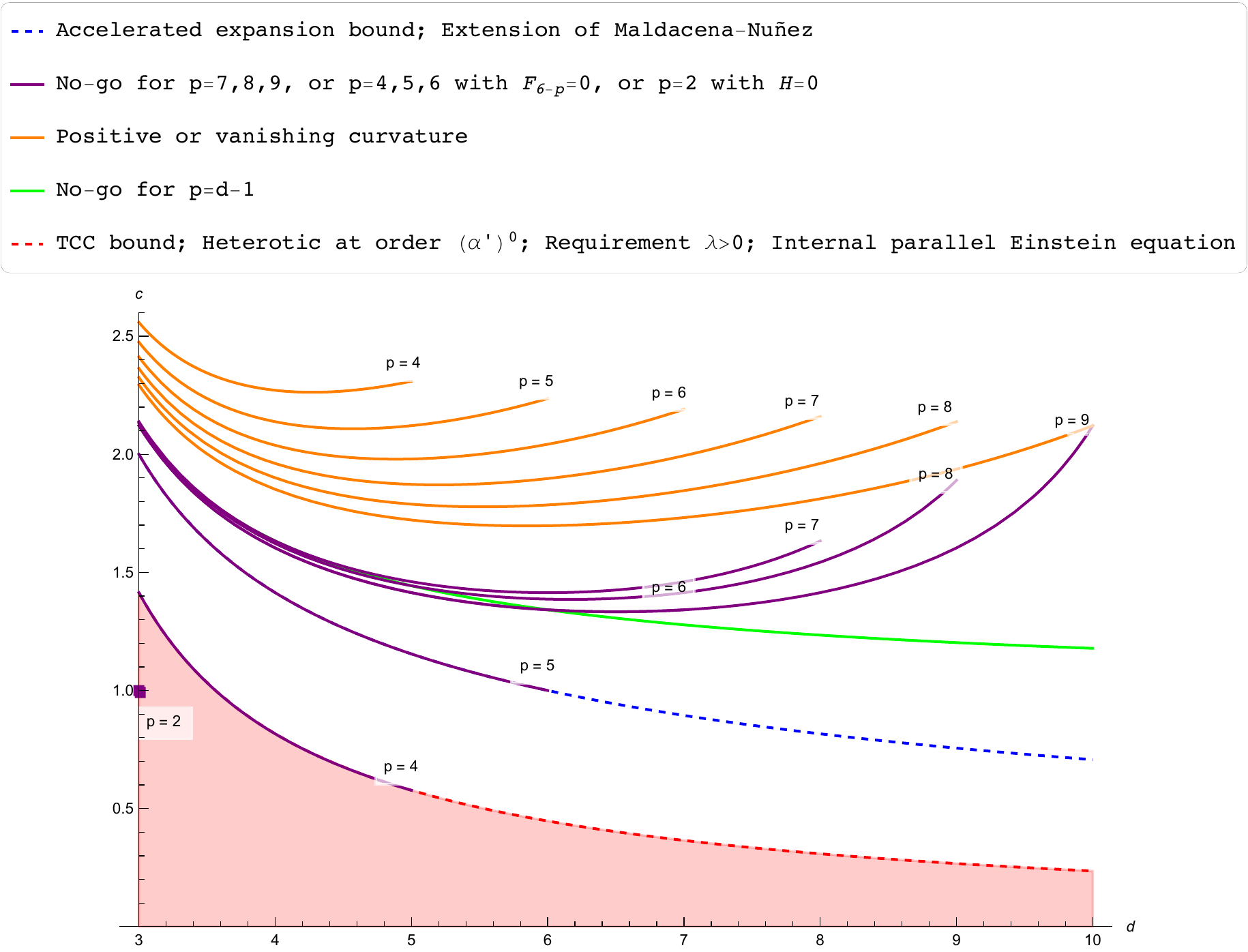}
\caption{Graph of the $c$ values in terms of the dimension $d$, $3 \leq d \leq 10$, for each no-go theorem indicated in Table \ref{tab:cvalues}. They are represented with plain lines of different colors, as made explicit in the caption above. Some values are $p$-dependent (dimensionality of $O_p/D_p$), as indicated on the graph. The dashed lines represent bounds: the TCC (lower) bound \eqref{TCC}, with its forbidden region being filled, and the so-called accelerated expansion (upper) bound, discussed in Section \ref{sec:accexp}. Interestingly, one no-go theorem violates the TCC bound, in $d=3$.}\label{fig:cvalues}
\end{figure}

The values for $c$ obtained in the 7 no-go theorems considered are summarized in Table \ref{tab:cvalues}. We also provide a graph of their dependency on the dimension $d$ in Figure \ref{fig:cvalues}. In both, we compare to the TCC bound.

\begin{table}[H]
  \begin{center}
  \noindent\makebox[\textwidth]{%
      \begin{tabular}{|c|c|c||c|}
    \hline
    & 10d & $d$-dim. & \\
No-go theorem & derivation & derivation & $c$ \\
 & (section) & (section) & \\
        \hhline{===::=}
    \multirow{2}{*}{Extension of Maldacena-Nuñez} & \multirow{2}{*}{\ref{sec:MN}} & \multirow{2}{*}{\ref{sec:MNd}} & \multirow{2}{*}{$\frac{2}{\sqrt{d-2}}$} \\
    &&&\\
    \hline
    \multirow{2}{*}{$p=7,8,9$, or $p=4,5,6$ \& $F_{6-p} =0$} & \multirow{4}{*}{\ref{sec:NonVanFlux}} & \multirow{4}{*}{\ref{sec:NonVanFluxd}} & \multirow{2}{*}{$\frac{2 (p-3)}{\sqrt{(d-2)(5d-1-4p-pd+p^2)}} \geq \color{red}{\frac{2}{\sqrt{(d-2)(d-1)}}}$} \\
    &&&\\
    \multirow{2}{*}{$p=2$ \& $H =0$ ($d=3$)} & & & \multirow{2}{*}{{\small ($1=$)}$\, \frac{2}{\sqrt{(d-2)(3d-5)}} < \color{red}{\frac{2}{\sqrt{(d-2)(d-1)}}} \, ${\small ($=\color{red}{\sqrt{2}}$)}} \\
    &&&\\
    \hline
    \multirow{2}{*}{Positive or vanishing ${\cal R}_{10-d}$} & \multirow{2}{*}{\ref{sec:PosCurv}} & \multirow{2}{*}{\ref{sec:PosCurvd}} & \multirow{2}{*}{$\frac{2(d+p-1)}{\sqrt{(d-2)(-1+d-dp+4p+p^2)}} > 1$} \\
    &&&\\
    \hline
    \multirow{2}{*}{$p=d-1$} & \multirow{2}{*}{\ref{sec:d-1}} & \multirow{2}{*}{\ref{sec:d-1d}} & \multirow{2}{*}{$\frac{d}{\sqrt{(d-2)(d-1)}}$} \\
    &&&\\
    \hline
    \multirow{2}{*}{Heterotic at order $(\alpha')^0$} & \multirow{2}{*}{\ref{sec:Het}} & \multirow{2}{*}{\ref{sec:Hetd}} & \multirow{2}{*}{\textcolor{red}{$\frac{2}{\sqrt{(d-2)(d-1)}}$}} \\
    &&&\\
    \hline
    \multirow{2}{*}{Requirement $\lambda > 0$} & & \multirow{2}{*}{\ref{sec:lambda}} & \multirow{2}{*}{\textcolor{red}{$\frac{2}{\sqrt{(d-2)(d-1)}}$}} \\
    &&&\\
    \hline
    \multirow{2}{*}{Internal parallel Einstein equation} & & \multirow{2}{*}{\ref{sec:InternEinstein}} & \multirow{2}{*}{\textcolor{red}{$\frac{2}{\sqrt{(d-2)(d-1)}}$}} \\
    &&&\\
    \hline
    \end{tabular}
    }
     \caption{No-go theorems considered in the paper, with the sections where they were derived, using the 10d approach or the $d$-dimensional one. The latter, giving an inequality of the form \eqref{eq:NoGoEq}, leads to a ($d$-dependent) $c$ value, reported in the last column. The values corresponding to the TCC bound \eqref{TCC} are highlighted in red.}\label{tab:cvalues}
  \end{center}
\end{table}

A remarkable result is that for $4 \leq d \leq 10$, all $c$ values obtained verify the TCC (lower) bound: $c \geq c_0$. There are several instances of saturation $c =c_0$ that turn out to be exactly those with saturation in $d=4$ \cite{Andriot:2020lea}. Let us emphasize that there is a priori {\sl no reason} for this to be the case. We only performed here a mere supergravity analysis, which is a priori totally unrelated to the TCC argument, that relies on a quantum gravity and cosmological principle. From the swampland perspective however, if one considers the TCC to be true, the two should match, since supergravity is a quantum gravity theory in an asymptotic limit, the classical and perturbative limit of string theory. From this point of view, the values obtained here provide a highly non-trivial check of the TCC proposal. Note in addition that while we observed already a saturation of the TCC bound in $d=4$ \cite{Andriot:2020lea}, giving $c = \sqrt{\frac{2}{3}}$, this could have been extended to any formula in $d$ dimensions (see Section \ref{sec:swampconj}): it is non-trivial to us that it is always extended to the TCC formula, setting the latter on solid grounds.\\

Another interesting result is what happens for $d=3$. We noticed several peculiarities for this dimension in no-go theorems, as indicated in \eqref{eq:p2nogo} and \eqref{eq:p2F0nogo}. It turns out that the (newly derived) no-go theorem \eqref{eq:p2nogo} or \eqref{cd=3p=2} gives a $c$ value that violates the TCC bound for $d=3$: $c < c_0$. It is the only instance where this happens. How should we interpret this violation of the bound?

Let us first examine the TCC \cite{Bedroya:2019snp}. The technical derivation of the bound does not suffer any issue in $d=3$. One may however question the physics starting point. Indeed, the physics argument relies on avoiding quantum fluctuations becoming classical. However, it is well-known that gravity in $d=3$  is very special: it is said to be topological, meaning that gravitational fluctuations are not possible. One could then argue that the TCC reasoning cannot be pursued any further since there is no fluctuation to consider and consequently, there is no reason to consider this bound in $d=3$. It would still be {\sl remarkable} that 10d supergravity is sensitive to this line of thought, and provides accordingly a counter-example to this bound. In addition, one may wonder whether the TCC physics reasoning could not be based on the scalar field fluctuations, instead of the (forbidden) gravitational fluctuations, although those two may eventually be related.

This discussion relates to a more general one, regarding the validity of swampland conjectures in $d=3$. While they are thought to hold true in $d \geq 4$, and not apply in $d \leq 2$ (see for instance all the $d-2$ factors encountered in this paper), the case of $d=3$ is not settled. Apart from the topological nature of gravity in $d=3$ with the consequences mentioned, one should also state that black holes are very different. Therefore, the standard gravity-based arguments motivating the swampland conjectures can become problematic in $d=3$, as suggested above for the TCC. This also gets reflected in the attempts for a weak gravity conjecture in $d=3$ (see e.g.~\cite{Montero:2016tif, Harlow:2022gzl}).

Some papers however successfully checked swampland conjectures in $d=3$, such as \cite{Alvarez-Garcia:2021hzz}. In \cite{Gonzalo:2018tpb, Gonzalo:2021zsp} some swampland conjectures are assumed to be valid in $d=3$ and then used, without reaching contradictions (so far). In those papers, one considers a simple circle dimensional reduction from $d=4$ to $d=3$, in which case one does not expect much to happen. We nonetheless note that a new ingredient could appear in lower dimensions that was not present in the initial theory, a situation probably not encountered through a simple dimensional reduction. We have witnessed such appearances and newly related complications in our study of de Sitter in various dimensions in Section \ref{sec:dSineachd}. It seems to be what happens here for the problematic no-go theorem and $c$ value: indeed, it is due to $O_2/D_2$ sources which can only appear in $d = 3$ and not in higher dimensions. As a result, they lead to new physics in lower dimensions, giving rise to an inconsistency with the TCC bound in $d=3$.

It would be interesting to find other concrete examples in $d=3$ that raise tensions with swampland conjectures, especially in the conjectures related to the TCC and de Sitter conjecture, to which we now turn. The type IIA scale-separated $AdS_3$ solutions of \cite{Farakos:2020phe, VanHemelryck:2022ynr} might be viewed as such examples.

\subsection{Related swampland conjectures}\label{sec:swampconj}

The check of the TCC bound \eqref{TCC}, as a minimal value for $c$, through supergravity no-go theorems across dimensions for $d \geq 4$, establishes this expression on solid grounds
\beq
M_p \, \frac{|\nabla V|}{V} \geq c \geq c_0 = \frac{2}{\sqrt{(d-2)(d-1)}} \ . \label{TCC2}
\eeq
As mentioned in Section \ref{sec:cvalues}, it is a non-trivial check of this expression. Even though a saturation had been observed in several cases in $d=4$, one could have imagined different expressions in other dimensions. One possibility comes from the analysis of the swampland distance conjecture in Calabi-Yau complex structure moduli space performed in \cite{Grimm:2018ohb} (see also \cite{Gendler:2020dfp}). Using the proposed relation to the de Sitter conjecture (see below), e.g.~taking $\alpha = \frac{1}{2}$ in \eqref{lc}, one would have obtained
\beq
c \geq \frac{4s}{\sqrt{10-d}} \ , \ \quad s=\begin{cases}1 & \text{if} \quad \frac{10-d}{2} \ \mbox{is even} \\
       \frac{1}{2} & \text{if} \quad \frac{10-d}{2} \ \mbox{is odd}\end{cases}
\eeq
where $\frac{10-d}{2}$ corresponds to the Calabi-Yau complex dimension. This is just one example of possible expressions for a lower bound or minimal value of $c$, that matches the $d=4$ one $c=\sqrt{\frac{2}{3}}$. We find it astonishing that we always get a single and same expression via the no-go theorems.

Another important proposal in the literature is that of \cite{Rudelius:2021oaz, Rudelius:2021azq}, revisited recently in \cite{Etheredge:2022opl, Rudelius:2022gbz}. As for the TCC, it is meant to hold in the asymptotics of field space. The proposed minimal value is higher than the TCC bound, and is given as follows
\beq
c \geq \frac{2}{\sqrt{d-2}} \ .\label{cRud}
\eeq
The main argument in favor of this value is that it remains invariant under dimensional reduction, as other swampland conjectures do. Interesting checks have also been performed, as well as a relation to the distance conjecture, which we will come back to.

As argued in \cite{Andriot:2020lea,Andriot:2021rdy}, the derivation of the $c$ value using supergravity no-go theorems is effectively single field. Indeed, the linear combination of derivatives entering the inequality \eqref{eq:NoGoEq} defines a (canonical) single field $\hat{t}_b$ as follows \cite{Andriot:2021rdy}
\beq
\sum_i \hat{b}_i\ \del_{\hat{\varphi}^i}  = \sqrt{\sum_i \hat{b}_i^2}\ \del_{\hat{t}_b} \ , \quad \quad a\, V + \sum_i M_p\, \hat{b}_{i}\, \del_{\hat{\varphi}^i}V \le 0  \ \Leftrightarrow\  \frac{c}{M_p}\, V + \del_{\hat{t}_b} V \le 0 \ .
\eeq
One then reproduces the inequality of the de Sitter conjecture using that $\del_{\hat{t}_b} V \geq - |\nabla V|$. A different way to view this is to consider the potential slope along one field direction only, while freezing all other fields to their (would-be) critical point value $\del_{\varphi^i} V =0$ or for us $\varphi^i=1$. Indeed, in that case, $|\del_{\hat{t}_b} V| = |\nabla V|$. This has been explicitly checked on the scalar potential $V(\tau,r)$ \eqref{rpot2} in $d=4$ in \cite[(4.31)]{Andriot:2020lea}: freezing one field leaves precisely the $e^{-c \, \hat{t}_b }$ asymptotic behaviour in the potential.

On the contrary, the bound \eqref{cRud} has been argued to hold when taking into account all fields, i.e.~the full gradient $|\nabla V|$, instead of freezing some \cite{Rudelius:2021oaz}. This allows a priori to get a higher, hence stronger, bound on $c$. This point of view seems to accommodate the TCC bound \eqref{TCC2} with the one in \eqref{cRud}. We hope to better understand in future work the distinction between these two bounds.\\

An important connection has been proposed between the swampland distance conjecture \cite{Ooguri:2006in, Klaewer:2016kiy, Baume:2016psm} and the de Sitter conjecture. It is proposed that the mass scale $m$ of the tower in the former is related to the scalar potential $V$ of the latter \cite{Ooguri:2018wrx, Bedroya:2019snp, Andriot:2020lea} in the asymptotics of field space
\beq
m \sim V^{\alpha} \label{mV}
\eeq
taking here $M_p=1$. This also relates to the anti-de Sitter distance conjecture \cite{Lust:2019zwm}, trading $V$ for $|\Lambda|$, since the asymptotics typically correspond to $|V| \rightarrow 0$ \cite{Dine:1985he}. The scalar potentials encountered are typically sums of exponentials. Considering the leading one in the asymptotics of field space, \eqref{mV} then relates the exponential rate $\lambda$ of the distance conjecture to the $c$ of the de Sitter conjecture: $\lambda = \alpha\, c$. This is in particular true for the lower bounds, i.e.~the minimal values of $\lambda$ and $c$; we refer to \cite{Andriot:2020lea} for a more detailed discussion. This leads to the proposed relation
\beq
\lambda_{{\rm min}} = \alpha\ c_{{\rm min}} \ . \label{lc}
\eeq
This shows that the value of $c_{{\rm min}}$ can also have an important impact on the distance conjecture. In \cite{Andriot:2020lea} it was proposed to take the TCC bound $c_{{\rm min}}=c_0$ and $\alpha = \frac{1}{2}$, given the good match in $d=4$ with values for $\lambda_{{\rm min}}$ appearing in the literature. The successful check of the TCC bound for $d \geq 4$ obtained in this paper motivates even more to consider
\beq
\lambda_{{\rm min}} = \frac{2\alpha}{\sqrt{(d-2)(d-1)}} \ ,
\eeq
with $\alpha(d=4) = \frac{1}{2}$, and possibly for any $d$. Let us also mention the possible range
\beq
\frac{1}{d} \leq  \alpha \leq \frac{1}{2} \ ,
\eeq
for (quasi-) de Sitter, proposed and argued in \cite{Rudelius:2021oaz, Castellano:2021mmx, Montero:2022prj}.

In \cite{Etheredge:2022opl}, a different interpretation of the above was considered by introducing a lower bound $\lambda_{{\rm min}}= \frac{1}{\sqrt{d-2}}$ that would correspond to the minimal rate of the {\sl lightest} tower. In other words, even though it is possible to have a tower decaying at a rate $\frac{1}{\sqrt{(d-2)(d-1)}}$, \cite{Etheredge:2022opl} suggests that one can then find a lighter tower, thus decaying with a higher rate, which has to be bigger than $\frac{1}{\sqrt{d-2}}$. This interpretation allows to verify a relation \eqref{lc} with $\alpha = \frac{1}{2}$, considering the bound \eqref{cRud} for $c$ that we discussed above. We hope to come back in future work to these intricate relations between conjectures, associated decay rates and bounds.

\subsection{Comments on the ``accelerated expansion bound''}\label{sec:accexp}

The bounds on $\frac{|\nabla V|}{V}$ discussed above are crucial for cosmology: not only do they drastically limit the possibility of having a quasi-de Sitter spacetime, they may even forbid accelerated expansion. Indeed, let us consider a $d$-dimensional cosmological model for $d\geq 3$, with a spacetime described by a FLRW metric, and a single (canonically normalized) scalar field $\phi$ with potential $V > 0$. Then, the following upper bound
\beq
\frac{|V'|}{V} < \frac{2}{\sqrt{d-2}} \ , \label{accexpbound}
\eeq
i.e.~$c = \frac{2}{\sqrt{d-2}}$ with the reduced Planck mass $M_p=1$, appears to be an asymptotic requirement for accelerated expansion (see e.g.~\cite{Obied:2018sgi,Rudelius:2021oaz,Rudelius:2022gbz}). To understand better this claim, we revisit here this bound and provide, in Appendix \ref{ap:accexp}, two distinct derivations of \eqref{accexpbound}.

As we see in Appendix \ref{ap:accexp}, having accelerated expansion implies that the kinetic energy should not be too large. Physically, we understand that this implies an upper bound on the slope $|V'|$, as in \eqref{accexpbound}. But details matter, and each of the two derivations of this bound rely on important assumptions that should not be neglected. The first one assumes slow-roll, and the second one considers an asymptotic limit to a fixed point. Whether the two assumptions are the same is not entirely clear to us, as discussed in Appendix \ref{ap:accexp}. In any case, this indicates {\sl ways to violate the bound \eqref{accexpbound} while still having accelerated expansion}: a first option would be to leave the slow-roll regime, and a second option would be to have a transient, i.e.~non-asymptotic, accelerated expansion phase (see related discussions in \cite{Cicoli:2021fsd}). These two options are worth being kept in mind, if one wants to avoid a cosmology of decelerated expansion. If we stick to the TCC bound \eqref{TCC2}, the so-called accelerated expansion bound \eqref{accexpbound} only leaves a small window
\beq
\frac{2}{\sqrt{(d-2)(d-1)}} \leq \frac{|V'|}{V} < \frac{2}{\sqrt{d-2}} \ ,
\eeq
depicted in Figure \ref{fig:cvalues}. Violating this upper bound with one of the two above options would then leave more room for e.g.~viable quintessence-like scenarios. We hope to come back to such interesting cosmological scenarios in future work.

\vfill

\subsection*{Acknowledgements}

We warmly thank N.~Cribiori, E.~Gonzalo and I.~Valenzuela for helpful exchanges related to this project. L.~H.~acknowledges support from the Austrian Science Fund (FWF): project number P34562-N, doctoral program W1252-N27.

\newpage

\begin{appendix}

\section{Derivations of the ``accelerated expansion bound'' in $d$ dimensions}\label{ap:accexp}

In this appendix, we provide two different derivations of the bound \eqref{accexpbound}, claimed to correspond to an asymptotic upper bound required for accelerated expansion. As discussed in Section \ref{sec:accexp}, we show that such an ``accelerated expansion bound'' holds upon assumptions, that one may consider violating.

We consider a $d$-dimensional spacetime, $d\geq 3$, described by a FLRW metric, and a (single) canonically normalized scalar field $\phi$ with a potential $V > 0$. The reduced Planck mass is here taken as $M_p=1$. As in a standard cosmological model, we start with the two Friedmann equations in $d$ dimensions and the equation of motion (e.o.m.) of $\phi$
\bea
& \frac{(d-1)(d-2)}{2} \left( H^2 + \frac{k}{a^2} \right) = \rho \ ,\\
& (d-2) \frac{\ddot{a}}{a} + \frac{d-3}{d-1} \rho + p =0  \ \Leftrightarrow \ \dot{H} - \frac{k}{a^2} + \frac{\rho + p}{d-2} = 0 \ , \\
& \ddot{\phi} + (d-1) H \dot{\phi} + V' = 0 \ ,
\eea
where we recall that $H= \frac{\dot{a}}{a}$, with a dot standing for a derivative with respect to the physical time $t$, and we took a homogenous scalar field. In the following, we will restrict ourselves to $k=0$. The scalar field energy density and pressure are given by
\beq
\rho = \frac{1}{2} \dot{\phi}^2 + V \ ,\quad p = \frac{1}{2} \dot{\phi}^2 - V \ ,
\eeq
and the equation of state parameter is given by $w = \frac{p}{\rho}$. Having an accelerated expansion requires, on top of $H >0$, to have $\ddot{a} >0$: from the second Friedmann equation, this amounts to having
\beq
\text{Accelerated expansion:}\quad w < -\frac{d-3}{d-1} \ . \label{accexpcond}
\eeq
This condition can be translated into the bound \eqref{accexpbound} in two ways, as we now show. Physically, one way to understand the bound is the following: the condition \eqref{accexpcond} indicates that the kinetic energy cannot be too important. The latter would be violated for a high potential slope, hence the upper bound on $|V'|$ in \eqref{accexpbound}.

\begin{itemize}
  \item {\bf Slow-roll approximation}

To impose proper slow-roll conditions in $d$ dimensions, let us first introduce the parameters
\beq
\epsilon \equiv - \frac{\dot{H}}{H^2} = \frac{1}{d-2} \frac{\dot{\phi}^2}{H^2} \ , \quad \eta \equiv \frac{\dot{\epsilon}}{H \epsilon} = 2 \frac{\ddot{\phi}}{\dot{\phi}H} + 2 \epsilon \ .
\eeq
The slow-roll approximation then amounts to
\beq
\epsilon \ll 1 \ ,\ \eta \ll 1 \ \Leftrightarrow \ \dot{\phi}^2 \ll H^2 (d-2) \ , \ \ddot{\phi} \ll \dot{\phi}H \ .
\eeq
This leads to the following simplified first Friedmann equation and $\phi$ e.o.m.
\beq
\frac{(d-1)(d-2)}{2}  H^2 = V \ ,\quad (d-1) H \dot{\phi} = - V'  \ , \label{simpplifslowroll}
\eeq
and one gets in turn that
\beq
\epsilon = \frac{d-1}{2} \frac{\dot{\phi}^2}{V} = \frac{d-2}{4}\, \frac{|V'|^2}{V^2} \equiv \epsilon_V \ ,\label{eps2}
\eeq
where we promoted the latter to the definition of $\epsilon_V$ in $d$ dimensions; we recover in this way the well-known slow-roll condition $\epsilon_V \ll 1$. Using the first expression for $\epsilon$ in \eqref{eps2}, it is then straightforward to expand $w$ in the slow-roll approximation as
\beq
w = -1 + \frac{2 \epsilon}{d-1} + {\cal O}(\epsilon^2) = -1 + \frac{d-2}{2(d-1)}\, \frac{|V'|^2}{V^2} + {\cal O}(\epsilon^2) \ .
\eeq
At first order in this slow-roll approximation expansion, we then recover from the accelerated expansion condition \eqref{accexpcond} the bound \eqref{accexpbound}
\beq
\frac{|V'|}{V} < \frac{2}{\sqrt{d-2}} \ . \label{accexpboundslow}
\eeq
This reproduces the bound on accelerated expansion $\frac{|V'|}{V} \le \sqrt{2}$ discussed in \cite{Halliwell:1986ja, Kitada:1991ih, Kitada:1992uh} for exponential potentials in $d=4$.

  \item {\bf Flow and asymptotic fixed points}

{\bf Note:} {\it During completion of this work appeared the paper \cite{Rudelius:2022gbz}. The following derivation is provided in the appendix of the latter. A minor difference is that our derivation considers generic potentials, while the derivation of \cite{Rudelius:2022gbz} and the original works in $d=4$ restrict to exponential potentials.}\\

We now take a different route and generalize the derivation of \cite{Copeland:1997et, Tsujikawa:2013fta} done in $d=4$ to any dimension $d\geq 3$. We introduce the variables
\beq
x=\frac{\dot{\phi}}{\sqrt{6}\, H} \ , \quad y=\frac{\sqrt{V(\phi)}}{\sqrt{3}\, H} \ ,
\eeq
for $V>0$. The original works include, on top of the above scalar field and potential, a matter energy density and pressure $\rho_m, p_m$. Taking the derivatives of $x$ and $y$ with respect to $N=\ln a$ results in
\bea
\frac{\d x}{\d N} &= - \sqrt{\frac{3}{2}} \frac{V'}{V} y^2 -(d-1)x + \frac{x}{d-2} \left( 6 x^2 + \frac{\rho_m + p_m}{H^2}  \right)  \label{eq:dx/dN}\\
\frac{\d y}{\d N} &= \sqrt{\frac{3}{2}} \frac{V'}{V} x y + \frac{y}{d-2} \left( 6 x^2 + \frac{\rho_m + p_m}{H^2}  \right)  \label{eq:dy/dN}
\eea
using the second Friedmann equation and the scalar e.o.m. The contributions $\rho_m + p_m$ can further be rewritten using the first Friedmann equation, as well as a parameter $w_m = \frac{p_m}{\rho_m}$. In the following, we rather set $\rho_m = p_m = 0$ to fit our initial setting.

We then determine the fixed points by setting $\frac{\d x}{\d N}=0,\ \frac{\d y}{\d N}=0$. Those correspond to the asymptotics of the $N$-flow, which is nothing but a time flow here. The value of $\frac{V'}{V}$ which then appears should be viewed as the asymptotic one (constant along the flow in the case of an exponential potential). The fixed points, and corresponding equation of state parameter $w$ (deduced from $(x,y)$) are then
\begin{enumerate}[label=(\alph*)]
\item $(x,y)=(0,0)\ , \quad \text{(for $V' \neq 0$), $w$ undetermined}$
\item $(x,y)=\left(\pm \sqrt{\frac{(d-1)(d-2)}{6}},0\right)\ , \quad w=1$
\item $(x,y)=\left(-\frac{(d-2)}{2 \sqrt{6}} \frac{V'}{V}, \pm \frac{\sqrt{(d-2)}}{2 \sqrt{6}} \sqrt{4(d-1)-(d-2) \left( \frac{V'}{V} \right)^2 } \right)  \ , \quad w=-1 + \frac{d-2}{2(d-1)} \left( \frac{V'}{V} \right)^2$
\end{enumerate}
Only the last fixed point allows for $V>0$, and is able to realise cosmic acceleration. Using its value for $w$ and the accelerated expansion condition \eqref{accexpcond}, we deduce the ``accelerated expansion bound'' \eqref{accexpbound}, as done already with slow-roll in \eqref{accexpboundslow}.\\
\end{itemize}

We see that in both derivations of the ``accelerated expansion bound'' \eqref{accexpbound}, there is a non-trivial assumption: the first one is slow-roll and the second one is an asymptotic limit to a fixed point. It is unclear to us that the two assumptions are the same. Still, let us note that the value of $x$ at the last fixed point is compatible with the two simplified equations \eqref{simpplifslowroll} obtained in the slow-roll approximation; also, $\frac{\d x}{\d N}$ is proportional to the parameter $\eta$. In any case, violating these assumptions could lead to a violation of the bound \eqref{accexpbound} while still having accelerated expansion, as we further discuss in Section \ref{sec:accexp}.

\end{appendix}

\newpage

\providecommand{\href}[2]{#2}\begingroup\raggedright
\endgroup

\end{document}